\documentclass[aps,twocolumn,superscriptaddress,preprintnumbers]{revtex4}
\usepackage{braket}
\usepackage{subfigure}
\usepackage[latin9]{inputenc}
\usepackage{color}
\usepackage{verbatim}
\usepackage{amsmath}
\usepackage{amssymb}
\usepackage{graphicx}
\usepackage{esint}
\usepackage{mathrsfs}
\usepackage{hyperref}

\makeatletter
\@ifundefined{textcolor}{}
{%
 \definecolor{BLACK}{gray}{0}
 \definecolor{WHITE}{gray}{1}
 \definecolor{RED}{rgb}{1,0,0}
 \definecolor{GREEN}{rgb}{0,1,0}
 \definecolor{BLUE}{rgb}{0,0,1}
 \definecolor{CYAN}{cmyk}{1,0,0,0}
 \definecolor{MAGENTA}{cmyk}{0,1,0,0}
 \definecolor{YELLOW}{cmyk}{0,0,1,0}
}

\usepackage{mathrsfs}

\draft
\makeatother

\begin{document}
\title{Magnetic effects on Chern Kondo insulator}

\author{Sen Niu}
\affiliation{International Center for Quantum Materials and School of Physics, Peking University, Beijing 100871, China}
\affiliation{Collaborative Innovation Center of Quantum Matter, Beijing 100871, China}
\author{Xiong-Jun Liu \footnote{Corresponding author: xiongjunliu@pku.edu.cn}}
\affiliation{International Center for Quantum Materials and School of Physics, Peking University, Beijing 100871, China}
\affiliation{Collaborative Innovation Center of Quantum Matter, Beijing 100871, China}

\begin{abstract}
We examine the Chern Kondo insulator proposed in a square optical lattice with staggered flux induced by $s$-$p$ orbital hybridization by revisiting its realization and taking into account the magnetic effects for the Kondo phases. The Ruderman-Kittel-Kasuya-Yoshida interaction is analyzed at the weak $s$-$p$ hybridization regime, with the anisotropic magnetic effects being discussed. Furthermore, the paramagnetic and magnetic phases coexisting with Kondo couplings are systematically investigated through the slave-boson theory, for which the rich phases are obtained, including the antiferromagnetic, collinear antiferromagnetic Kondo insulator, and Kondo metal phases. The magnetic orders are shown to enhance the effective Kondo hybridization compared with the case without taking into account magnetic effects, and exhibit different influences on the bulk topology. In particular, the antiferromagnetic ordering always enhances the topological phase by increasing bulk gap of the Chern Kondo phases. The results show the rich topological and magnetic effects obtained in the present Chern Kondo lattice model. We also investigate how to identify the topology and strong correlation effects through measuring the Hall conductance and double occupancy, which are achievable in ultracold atom experiments.
\end{abstract}

\date{\today }
\maketitle
 \section{Introduction}

In the past decade the topological states of quantum matter have been extensively studied in condensed matter physics, such as the time-reversal invariant topological insulators (TIs) with uncorrelated gapped bulk band and gapless surface states~\cite{TI1, TI2}. Recently a class of strongly correlated topological phase have been predicted in heavy fermion systems, called topological Kondo insulators (TKIs)~\cite{TKI1, TKI2,TKI3}. These heavy fermion systems usually have itinerant $d$ orbitals and localized $f$ orbitals with strong coulomb interactions. With the $d$-$f$ hybridization, a narrow gap due to the formation of Kondo singlets which screen the local moments would develop at low temperature. The theoretically proposed TKIs have been supported by the transport measurement~\cite{conduct1, conduct2,conduct3}, photo emission~\cite{arpes1, arpes2, arpes3, arpes4, arpes5} and scanning tunneling spectroscopy~\cite{stm1, stm2}.

While the effective band description for strongly correlated TKIs seems indistinguishable from that of an uncorrelated TI~\cite{correlation1}, new effects arising from electronic correlations in TKIs have been predicted in Refs.~\cite{correlation1,correlation2,correlation3,correlation4,correlation5,correlation6,correlation7,correlation8,correlation9}. Moreover, the observation of bulk quantum oscillations~\cite{osc1,osc2}, linear specific heat, anomalous thermal, and optical conductivity~\cite{thermal1,thermal2,thermal3,thermal4} have raised the possibility of a neutral Fermi surface in the bulk of the TKI $\text{SmB}_6$~\cite{correlation6}. On the other hand, the recent rapidly developing new technologies for ultracold atoms which are clean and fully controllable may provide novel opportunities for studying many-body physics and topological phases, see e.g. the Refs.~\cite{coldatom1,coldatom2,coldatom3,coldatom4,coldatom5,coldatom6,coldatom7,coldatom8,coldatom9,coldatom10,coldatom11,coldatom12,coldatom13,coldatom14,coldatom15,coldatom16,dynamicaltopo1,dynamicaltopo2}. In this paper we discuss one such example, a strongly correlated quantum anomalous Hall (QAH) phase, called Chern Kondo (CK) insulator which was proposed recently in an optical lattice~\cite{chernkondo}.

The original idea of CK insulator~\cite{chernkondo} is summarized as following. Consider a checkerboard superlattice with $s$ orbitals on A  sites and $p_X$ orbitals on B sites. Due to anistropy of the superlattice, the nearest neighbour hopping between $p_X$ orbitals is along $\hat{X}$  direction and forms an itinerant $p_X$ band, while the nearest neighbour hopping between $s$ orbitals is along $\hat{Y}$ direction, forming a nearly flat band and lying below $p_X$ band. Through optical assisted Feshbach resonance~\cite{feshbach1,feshbach2,feshbach3,feshbach4,feshbach5,feshbach6,feshbach7} the repulsive on-site interaction for $s$ orbitals is tuned to be strong, while the on-site interaction for $p_X$ orbitals is negligible. Without $s$-$p_X$ hybridization, the $s$ orbitals on A sites form a Mott insulator at half filling. By laser assisted tunneling~\cite{gaugefield1,gaugefield2,gaugefield3,gaugefield4}, the $s$-$p_X$ hybridization is induced and a periodic Anderson model with laser-induced staggered flux is realized. When the hybridization is tuned to exceed a critical value, the Kondo phase emerges with a finite $s$-$p_X$ hybridization gap being formed. The gapped quasiparticle band results in a non-trivial correlated Chern insulator with QAH effect~\cite{chernkondo}. Difference between the noninteracting Chern insulator and CK insulator can be detected by measuring the band topology and double occupancy experimentally.

Nevertheless, there are important issues of the CK insulating phase which were not well addressed in the previous work~\cite{chernkondo}. First of all, we examine in detail the realization of the Chern insulating phase, and found that the previous scheme is not applicable to generate a staggered flux pattern for the checkerboard lattice, which is essential to realize the Chern insulating phase. Secondly, in the original work, only paramagnetic state was considered, while it was shown that in Kondo lattice problems the Ruderman-Kittel-Kasuya-Yoshida (RKKY) interaction between the localized electrons competes with the Kondo effect, resulting in magnetic phases in weak coupling regime~\cite{doniach}, and in periodic Anderson lattice problems the magnetic instability also occurs in the Kondo phases~\cite{anderson,sbmf,dorin,dmrg}. It is then important to investigate the possible existence of magnetic phases in the CK model, and the effect of magnetism on the CK phase and the phase transition. In this work we fully address these issues and uncover nontrivial topological Kondo physics which were not predicted previously. In particular, we improve the previous realization and propose a new feasible scheme to generate the $s$-$p_X$ orbital hybridization with a staggered flux which can induce QAH phase in noninteracting regime. Moreover, we study systematically the magnetic effects on the strongly correlated QAH phase based on RKKY interaction and the slave-boson mean-field theory. We map out the magnetic and QAH phase diagram, and show that the magnetic orders depend on the $s$-$p_X$ hybridization strength and the flux $\phi_0$ generated by the laser assisted tunneling. The different magnetic orders can have different influences on QAH phases. Interestingly, a significant enhancement of the correlated QAH effect by the magnetic ordering is predicted.

The structure of this paper is organized as follows. In Sec. II we examine the realization of the CK model in detail, and propose a new feasible scheme for the realization based on the previous one~\cite{chernkondo}. In Sec. III, we derive the effective Kondo lattice Hamiltonian and RKKY interaction, with which we investigate the magnetic effects controlled by $s$-$p_X$ hybridization. Sec. IV presents a systematically study of the ground state magnetism and QAH phase diagram through the slave-boson theory. Especially, in this section we also study the influences of topology and strong correlation on the CK phase, which can be identified by measuring Hall conductance and double occupancy in cold atom experiments. The conclusions are given in the last Sec. V.

 \section{The improved scheme for realization}

In the realization of CK insulator~\cite{chernkondo}, the laser assisted tunneling is applied to generate the complex $s$-$p_X$ hybridization which is associated with a staggered magnetic flux. Nevertheless, in subsection A we point out that the original laser-assisted tunneling failed to create the required flux, and a modified configuration is necessary. In subsection B we improve the original method and propose a feasible scheme for the realization, as shown in [Fig.~\ref{Jxy}] following the method in Ref.~\cite{chiraltopologicalorder}. Different from the previous proposal which applies a beam running in the $x-y$ plane to induce Raman transition~\cite{chernkondo}, in the present new scheme the Raman beam propagates along $\hat{z}$ direction and has a phase difference between $\hat{x}$ and $\hat{y}$ polarization components. We show that the staggered flux of a minimal plaquette can be tuned freely from $0$ to $2\pi$, as required for realizing the CK insulator.

\begin{figure}[t]
\centering
\includegraphics[width=3in]{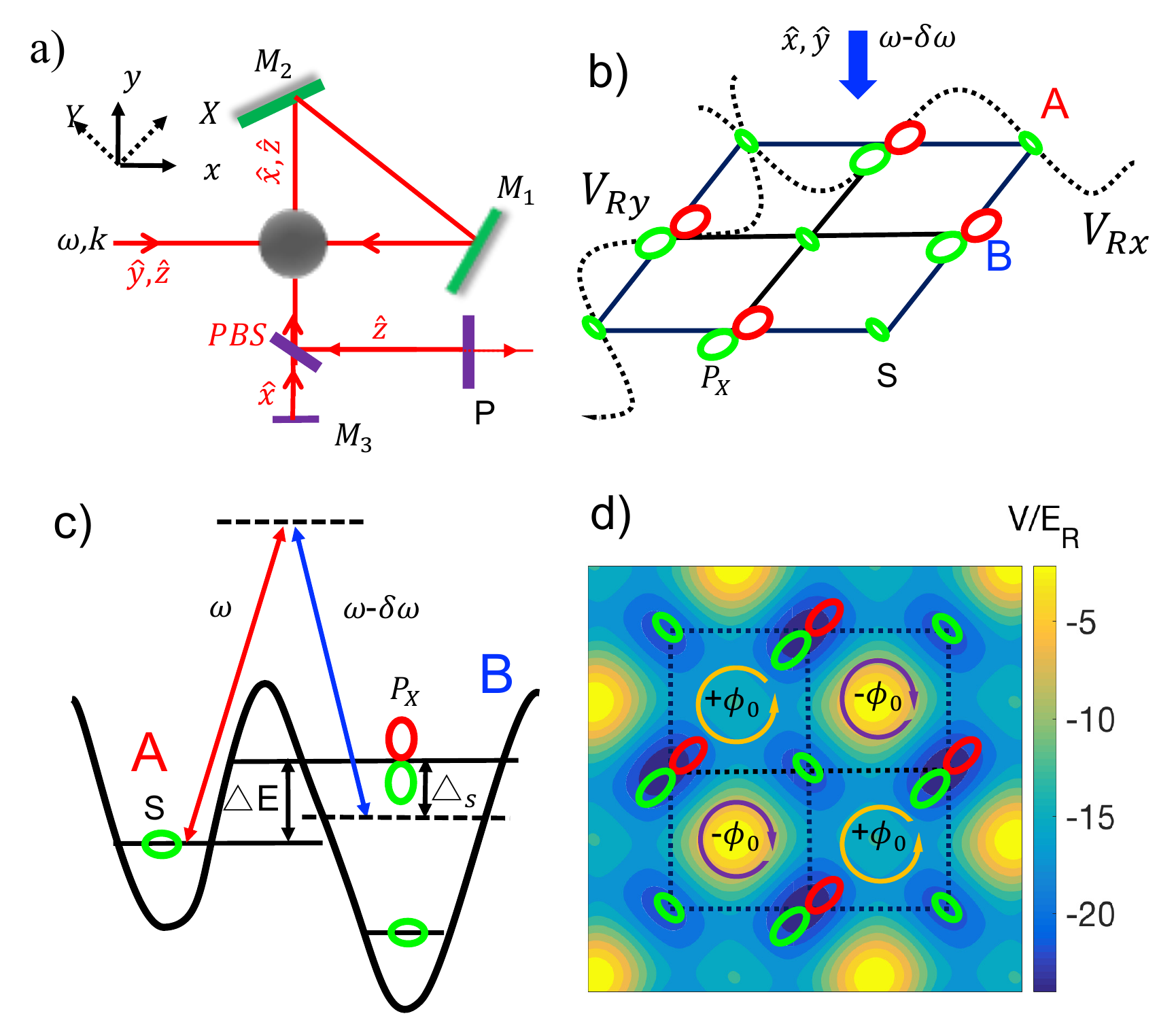}
\caption{a) Sketch of generating double-well checkerboard superlattice by a laser beam proposed in Ref.~\cite{chernkondo}. b) The checkerboard sublattice and Raman potential for the improved scheme. c)-d) The Raman coupling that induces $s$-$p_X$ orbital hybridization (c)) and creates staggered flux pattern displayed in d), with the lattice potential parameters taken as $(V_0,V_1,V_2)=(8,8,5)E_R$ and $\phi_0=2\phi$.}
\label{setup}
 \end{figure}

 \subsection{Synthetic flux: the previous model}

\begin{figure}[t]
\centering
\includegraphics[width=3.4in]{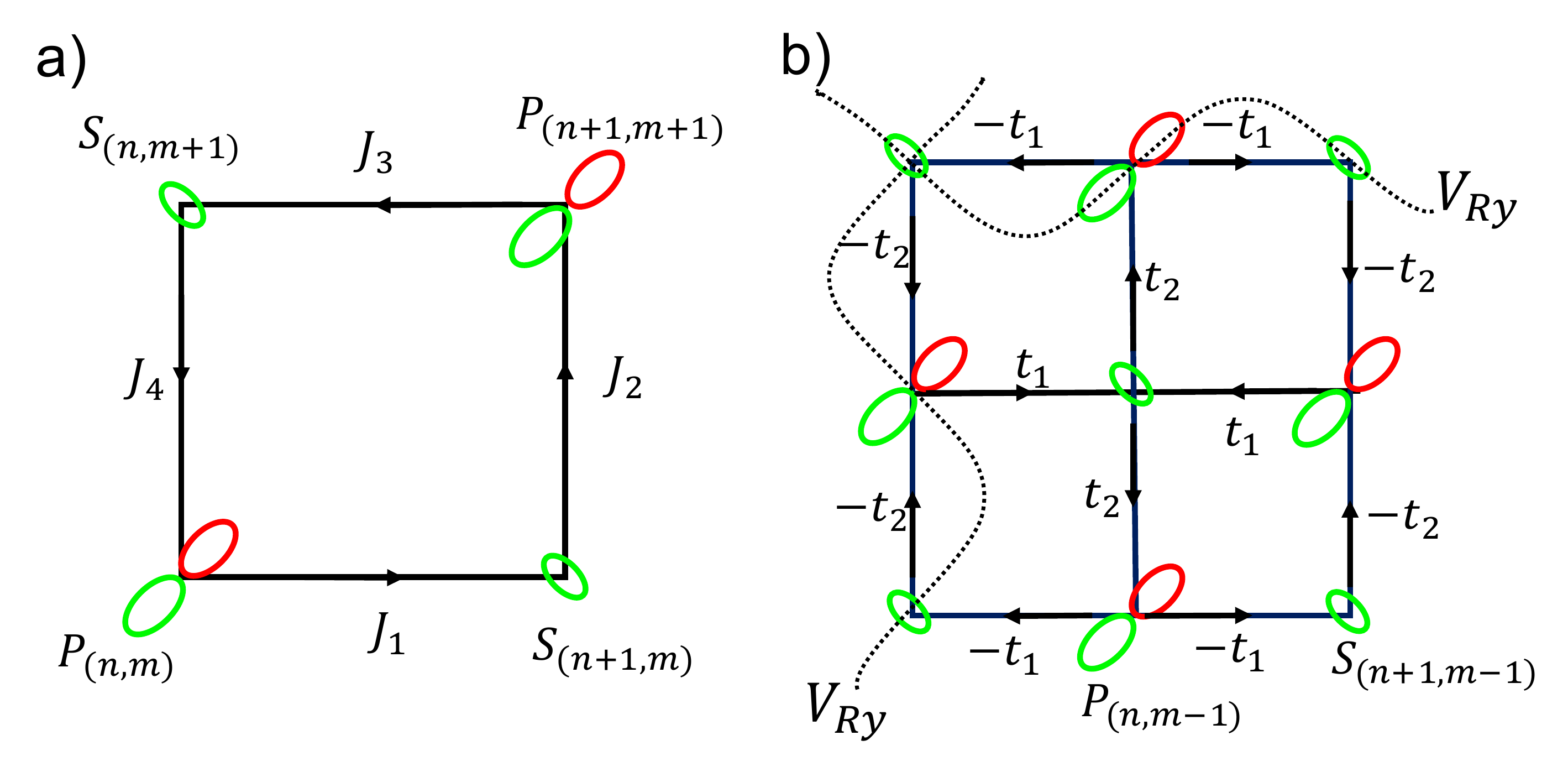}
\caption{a) Laser induced $s$-$p_X$ hoppings in previous setup~\cite{chernkondo} for one plaquette of the double-well optical lattice. b) Laser induced $s$-$p_X$ hybridization in the new scheme for four nearest plaquette of the double-well optical lattice.}
\label{tunnel}
 \end{figure}

 In the previous proposal~\cite{chernkondo}, the $s$-$p_X$ orbital hybridization is induced by an effective Raman potential $V_R=V_m\cos(\delta\omega t+k_Ry)$, where $\delta\omega$ and $k_R$ are the frequency and wave vector of the Raman potential, respectively. The $s$-$p_X$ orbital hybridization is induced when the frequency difference $\delta\omega$ compensates the energy difference between $s$ and $p_X$ orbitals, and can be calculated through the rotating-wave approximation. To examine the hopping and flux generated by the $V_R$, we calculate the hopping integrals of a loop round a minimal square (see Fig.~\ref{tunnel} a) and find that the hopping integrals take the following form (details of the calculation can be found in Appendix):
   \begin{equation}
  \begin{aligned}
&J_1=\int d^2r\psi_{n,m}^p\psi_{n+1,m}^se^{ik_Ry}=e^{ik_Rm}I_a,\\
&J_2=\int d^2r\psi_{n+1,m+1}^p\psi_{n+1,m}^se^{-ik_Ry}=-e^{-ik_R(m+\frac{1}{2})}I_b^*,\\
&J_3=\int d^2r\psi_{n+1,m+1}^p\psi_{n,m+1}^se^{ik_Ry}=-e^{ik_R(m+1)}I_a^*,\\
&J_4=\int d^2r\psi_{n,m}^p\psi_{n,m+1}^se^{-ik_Ry}=e^{-ik_R(m+\frac{1}{2})}I_b,\nonumber
  \end{aligned}
  \end{equation}
where $\psi_{n,m}^p$ ($\psi_{n,m}^s$) denotes real maximally localized Wannier function for $s$ ($p_X$) orbital localized at the site $(m,n)a'$, with $a'=a/\sqrt{2}$ and $a$ being the lattice constant of sublattice. For simplicity we set the lattice constant $a=1$. The orbitals $\psi_{0,0}^s(x,y)$ and $\psi_{0,0}^p(x,y)$ are parity even and parity odd, respectively. The quantities $I_a$ and $I_b$ are complex. From the above result it is apparent that the product of four hopping integrals is real, implying zero synthetic magnetic flux over a loop of a plaquette. Even though the flux over a triangular $(n,m)\rightarrow (n+1,m) \rightarrow (n+1,m+1)\rightarrow (n,m)$ might be non-zero. Thus time reversal symmetry is broken, while we can show that this does not lead to QAH effect. If we apply a gauge transformation $c_{s,n,m}^{\dagger} \rightarrow c_{s,n,m}^{\dagger}e^{ik_Rm}$, in the new basis the hopping integrals become
     \begin{equation}
  \begin{aligned}
&J_1=I_a,\\
&J_2=-e^{\frac{i}{2}k_R}I_b^*,\\
&J_3=-I_a^*,\\
&J_4=e^{-i\frac{i}{2}k_R}I_b.
  \end{aligned}
  \end{equation}
From the above results one can easily find that the tight-binding Hamiltonian in k-space lacks the $\tau_x$ term, and cannot lead to QAH effect.

\subsection{The improved scheme}

Note that the optical lattice potentials for the checkerboard lattice (see Fig.~\ref{setup} a)) are formed by the following standing wave fields~\cite{chernkondo}
\begin{equation}
\begin{aligned}
&\bold E_{xy}=2E_1[\cos (k_0x)\hat{y}+\cos (k_0y)\hat{x}],\\
&\bold E_{z}=E_2[e^{ik_0x}+e^{ik_0y}+\alpha e^{-ik_0x+i\pi}+\alpha e^{ik_0y+i\pi}]\hat{z},
\end{aligned}
\end{equation}
which induces the optical lattice potential as $V_{\rm latt}=-V_0[\cos^2(k_0x)+\cos^2(k_0y)]-V_1\sin^2[(k_0/2)(x-y)]-V_2\sin^2[(k_0/2)(x+y)]$, with the parameter $\alpha$ used to tuning the relative magnitudes of the lattice depths $V_{0,1,2}$.

In the present improved scheme, to generate staggered synthetic magnetic flux, we add an additional incident beam with electric field
\begin{eqnarray}
\tilde {\bold E}_{xy}=E_m e^{ik_zz-i(\omega-\delta\omega)t}(e^{i\phi_x}\hat{x}+e^{i\phi_y}\hat{y}),
\end{eqnarray}
which propagates along $z$ direction and is polarized in the $x-y$ planes. The Raman potential see Fig.~\ref{setup} b) is generated by $\tilde {\bold E}_{xy}$ together with the optical lattice beam $\bold E_{xy}$. For the 2D system one can set that the 2D plane is located at $z=0$. The Raman potential takes the form
\begin{equation}
\begin{aligned}
  V_R\propto E_1E_m e^{i\delta\omega t}(\cos k_0ye^{i\phi_x}+\cos k_0xe^{i\phi_y}+\mathrm{H.c.}).
  \end{aligned}
\end{equation}

In the rotating-wave approximation, the effective A-B on-site energy difference becomes $\Delta_s=\Delta E-\delta \omega$, where $\Delta E$ is the bare energy difference between $p_X$ and $s$ orbital, as shown in [Fig.~\ref{setup} c)]. We now compute the hopping integrals generated by $V_R$. To illustrate the feature of the hopping integral, we consider four small plaquettes [Fig.~\ref{tunnel} b)], and calculate the hopping integrals by
\begin{equation}
\begin{aligned}
J_{(n,m)\rightarrow (n,m+1)}&=t_ae^{-i\phi_y}+t_be^{-i\phi_x},\\
J_{(n,m+1)\rightarrow (n-1,m+1)}&=-t_ae^{i\phi_x}-t_be^{i\phi_y},\\
J_{(n-1,m+1)\rightarrow (n-1,m)}&=-t_ae^{-i\phi_y}-t_be^{-i\phi_x},\\
J_{(n-1,m)\rightarrow (n,m)}&=t_ae^{i\phi_x}+t_be^{i\phi_y}.
\end{aligned}
\label{newfluxformuler}
\end{equation}
Here the coefficients $t_a$ and $t_b$ are real quantities, and are calculated by
\begin{equation}
\begin{aligned}
&t_a=\int d^2r\psi_{-1,0}^p(x,y)\psi_{0,0}^s(x,y)\cos(k_0y),\\
&t_b=-\int d^2r\psi_{-1,0}^p(x,y)\psi_{0,0}^s(x,y)\cos(k_0x).
\end{aligned}
\end{equation}

To obtain the phases of hopping integrals in Eq.~\eqref{newfluxformuler}, we define
 \begin{equation}
\begin{aligned}
&t_1=t_ae^{i\phi_x}+t_be^{i\phi_y},\\
&t_2=t_ae^{-i\phi_y}+t_be^{-i\phi_x},
\end{aligned}
\label{definehop}
\end{equation}
as shown in [Fig.~\ref{tunnel} b)]. The product of the four hopping integrals in Eq.~\eqref{newfluxformuler} equals to
 \begin{equation}
\begin{aligned}
(t_1 t_2)^2=(t_a^2e^{i\phi_x-i\phi_y}+t_b^2e^{i\phi_y-i\phi_x}+2t_at_b  )^2.
\end{aligned}
\label{product}
\end{equation}

It can be easily seen that when $\phi_x\neq\phi_y$ and $ t_a \neq t_b$, the above product is complex, leading to a nonzero staggered flux across a plaquette as illustrated in [Fig.~\ref{setup} d)] and [Fig.~\ref{tunnel} b)].

The magnitudes of $t_a$ and $t_b$ can be computed using the Maximally localized Wannier functions $\psi_p$ and $\psi_s$. On the other hand, in this work we shall consider the tight-binding regime, in which case, as a good approximation, the coefficients $t_{a,b}$ can be numerically calculated in the following approximate way. We take a rectangle piece of lattice potential containing a single $s$-$p_X$ double well and solve the orbital wavefunctions, which replace the Wannier functions in computing $t_{a,b}$. With the parameter condition that $(V_0,V_1,V_2)=(8,8,5)E_R$~\cite{chernkondo}, where $E_R=\hbar^2 k_0^2/(2m)$ is the recoil energy, we find $t_a\approx -2 t_b$. Now the phase $\phi_0$ of a plaquette in Eq.~\eqref{product} can be simplified to
 \begin{equation}
\begin{aligned}
\phi_0=2\phi=2\arctan{\frac{3\sin{(\phi_x-\phi_y)}}{5\cos{(\phi_x-\phi_y)}+4}},
\end{aligned}
\end{equation}
so the total flux $\phi_0$ of a plaquette can be tuned from $-\pi$ to $\pi$ through tuning the phase difference $\phi_x-\phi_y$.

From the relative configuration of lattice and Raman coupling potentials [Fig.~\ref{tunnel} b)], we can verify easily that $J_{(n-1,m)\rightarrow (n,m)}=-J_{(n,m)\rightarrow (n+1,m)}$. For simplicity, we perform the gauge transformation
\begin{equation}
\begin{aligned}
s_{m,n,\sigma}^{\dagger}\rightarrow (-1)^{m}s_{m,n,\sigma}^{\dagger},\\
p_{m,n,\sigma}^{\dagger}\rightarrow e^{-i\phi_1}p_{m,n,\sigma}^{\dagger},\\
\end{aligned}
\label{gauge}
\end{equation}
where $\phi_1$ is the phase of hopping integral $t_1$ determined through Eq.~\eqref{definehop}. With above gauge transformation the hopping integrals for $s$-orbitals reverses sign $t_s\rightarrow-t_s$, and the phase of $s$-$p_X$ hopping integral $t_1$ along $\hat{x}$ direction is transferred to the hopping integral $t_2$ along $\hat{y}$ direction so that $\pm t_1\rightarrow t_{sp}$ and $\pm t_2\rightarrow t_{sp}e^{i\phi}$. Finally the tight binding Hamiltonian $H=H_0+H_{int}$ reads

\begin{eqnarray}
H_0&=&\sum\limits_{i\sigma} \left[ t_{s}^Ys_{i\sigma}^{\dagger} s_{i\pm\hat{Y}\sigma}-\Delta_s    s_{i\sigma}^{\dagger} s_{i\sigma}+ t_{p}^Xp_{Xi\sigma}^{\dagger} p_{Xi\pm\hat{X}\sigma}\right]\nonumber\\
&&+\sum_{\langle ij\rangle \sigma}F(\bold r)s_{i\sigma}^{\dagger}p_{Xj\sigma}\delta_{j,i+\bold r}+\mathrm{H.c.},\\
H_{int}&=&U_s\sum\limits_{i}\hat{n}_{si\uparrow}\hat{n}_{si\downarrow},
\end{eqnarray}
where $s_{j\sigma}^{\dagger}$/$p_{Xj\sigma}^{\dagger}$ are creation operators for $s$/$p_X$ orbitals at the $j$-th site, $\sigma=\uparrow,\downarrow$, $F(\bold r)=t_{sp}$ for $\bold r=\pm \hat{x}$ and $F(\bold r)=t_{sp}e^{-i\phi}$ for $\bold r=\pm \hat{y}$ after gauge transformation Eq.~\eqref{gauge}. The interaction part is tuned by Feshbach resonance~\cite{chernkondo} and we only study the strong repulsive $U_s$ limit.
For convenience, we rotate the $\hat{x}-\hat{y}$ coordinate frame by $90^{\circ}$ to $\hat{X}-\hat{Y}$ coordinate frame where $\hat{X}=\hat{x}+\hat{y},\hat{Y}=-\hat{x}+\hat{y}$ when we Fourier transform the tight binding Hamiltonian. We will use this coordinate frame in the remaining parts. In the $\bold k$-space, the $s$ and $p_X$ orbital has dispersion relation $\epsilon_{s\bold k}=2t_s^Y\cos{k_Y}$ and $\epsilon_{p\bold k}=2t_p^X\cos{k_X}$ respectively. The single particle part of the tight binding Hamiltonian is written as $H_0=\sum_{\bold k\sigma}\mathcal{C}_{\bold k \sigma}^{\dagger}\mathcal{H}_0(\bold k)\mathcal{C}_{\bold k \sigma}$ with $\mathcal{C}_{\bold k \sigma}^{\dagger}=(s_{\bold k \sigma}^{\dagger},p_{X\bold k \sigma}^{\dagger})$ and the Bloch Hamiltonian takes the form
  \begin{equation}
\begin{aligned}
\mathcal{H}_0(\bold k)=d_0(\bold k)\tau_0+d_x(\bold k)\tau_x+d_y(\bold k)\tau_y+d_z(\bold k)\tau_z,
\end{aligned}
\end{equation}
   where $d_x(\bold k)=2t_{sp}(\cos{\phi}\cos{\frac{k_X+k_Y}{2}}+\cos{\frac{k_X-k_Y}{2}})$, $d_y(\bold k)=2t_{sp}\sin{\phi}\cos{\frac{k_X+k_Y}{2}}$, $d_{0/z}(\bold k)=\pm t_p^X\cos{k_X}+ t_s^Y\cos{k_Y}-\Delta _s/2$ and the Pauli matrix $\tau_{x,y,z}$ act on orbital space. The single-particle Hamiltonian $H_0$ leads to a QAH phase when $|\Delta_s|<2(t_s+t_p)$ and $0<\phi<\pi$~\cite{chernkondo}.

\section{RKKY magnetic interaction}

The RKKY interaction in Kondo lattice systems usually refers to the indirect coupling between local moments induced by the hybridization between local $f$ electrons and itinerant $d$ electrons in solid state physics. The competition between RKKY interaction (characterised by the N\'{e}el temperature $|J_K\rho|^2$) and Kondo effect (characterised by the Kondo temperature $T_K\sim \exp(-1/|J_K\rho|)$) is described by the Doniach diagram~\cite{doniach} which states that the RKKY interaction dominates in the weak $|J_K|$ limit and the Kondo effect dominates in the large $|J_K|$ limit. To investigate the possible magnetic phases in our CK model, we derive the effective RKKY interaction in this section through two steps. We first derive the effective Kondo lattice Hamiltonian from our CK model in subsection A, and then derive the effective RKKY interaction in subsection B based on the Kondo lattice Hamiltonian obtained in subsection A. We also analyse the static magnetic susceptibility of the RKKY interaction in subsection B. The static susceptibility is affected by the Fermi surface nesting effect and the $\phi$-dependent hybridization. We will show that the Fermi surface nesting effect of the $p_X$ band always favors staggered magnetic order in $\hat{X}$ direction, while the hybridization may favor different magnetic orders for different phase $\phi$ of the $s$-$p_X$ hybridization.

 \subsection{Effective Kondo lattice Hamiltonian}

The Kondo lattice Hamiltonian is the effective Hamiltonian derived from the periodic Anderson model by eliminating valence fluctuations and performing second-order perturbation in the Kondo regime where the hybridization is weak and the local orbital on-site energy lies far below the Fermi level of itinerant band. This step can be done either by Schrieffer-Wolff transformation~\cite{Schrieffer} or by projection operator~\cite{book}. In this paper, we take the latter method to handle our CK model, which is also convenient for deriving the RKKY interaction.

We perform the perturbation in the $t_s^Y=0$ and $U_s=+\infty$ limit. The Hamiltonian $H=H_1+H'$ is separated into two parts: $H_1$ preserves occupancy number of $s$ orbital atoms, and the hybridization term $H'$ mixes the subspaces with different number of $s$-orbital atoms:

 \begin{equation}
    \begin{aligned}
H_1=&\sum\limits_{i\sigma}\left[ -\Delta_ss_{i\sigma}^{\dagger}s_{i\sigma}+t_{p}^Xp_{Xi\sigma}^{\dagger} p_{Xi\pm \hat{X}\sigma} \right]\\
&+\sum\limits_{i}U_s\hat{n}_{si\uparrow}\hat{n}_{si\downarrow},\\
H'=&\sum\limits_{\bold k,i}\frac{V_{\bold k}e^{-i\bold k\cdot \bold R_i}}{\sqrt{N}}s_{i\sigma}^{\dagger}p_{Xk\sigma}+\mathrm{H.c.},
  \end{aligned}
  \end{equation}
where $V_{\bold k}=2t_{sp}[\exp{(i\phi)}\cos{\frac{k_X+k_Y}{2}}+\cos{\frac{k_X-k_Y}{2}}]$ is the hybridization function in $\bold k$ space, $N$ denotes the number of unit cells, and $\bold R_i$ denotes the $s$ orbital position in $\hat{X}$-$\hat{Y}$ coordinate frame. The $H'$ term is treated as a perturbation if the hybridization strength is weak. Experimentally, the magnitude of $t_{sp}$ can be tuned independently of the optical lattice by the strength of the Raman laser. To obtain the effective Hamiltonian, we define projection operator $P$ and $Q=1-P$, where $P$ projects onto the subspace in which each $s$ orbital is singly occupied. From the Schr\"{o}dinger equation, we obtain the following equations:
 \begin{equation}
    \begin{aligned}
&(P+Q)H(P+Q)\psi=E\psi,\\
&PH(P+Q)\psi=EP\psi,\\
&QH(P+Q)\psi=EQ\psi.
   \end{aligned}
  \end{equation}

We eliminate $Q\psi$ to obtain effective Hamiltonian in subspace $P$:
 \begin{equation}
    \begin{aligned}
H_p(E)=PHP-PHQ\frac{1}{QHQ-E} QHP,
   \end{aligned}
  \end{equation}
  where $H_p(E)$ is the effective Hamiltonian that satisfies
   \begin{equation}
    \begin{aligned}
    H_p(E)[P\psi]=E[P\psi].
   \end{aligned}
  \end{equation}

The approximation we then consider is to substitute the unknown eigenenergy $E$ by the energy of the unperturbed states $E_0$. The product of operators $PHQ$ and $QHP$ only keeps second order virtual processes that the states in subspace $P$ transfers to subspace $Q$ and then return back to subspace $P$. After some algebra (see the Appendix for details) we obtain the effective Kondo lattice Hamiltonian:
 \begin{equation}
    \begin{aligned}
H_{KL}= \sum\limits_{\bold k\sigma}\epsilon_{p\bold k} p_{X\bold k\sigma}^{\dagger}p_{X\bold k\sigma}+\sum\limits_{i,\bold k,\bold k'}2J_{\bold k,\bold k',i}\bold{S}_i\cdot \bold{s}_{\bold k \bold k'},
   \end{aligned}
  \end{equation}
where we have defined the $s$-orbital spin operators $\bold{S}_i=s_{i\sigma^{\prime}}^{\dagger}\boldsymbol \tau_{\sigma ' \sigma}s_{i\sigma}$, $p_X$-orbital spin operators $\bold{s}_{\bold k,\bold k'}=p_{X\bold k\sigma^{\prime}}^{\dagger}\boldsymbol \tau_{\sigma ' \sigma}p_{X\bold k'\sigma}$, and the anistropic $\bold k$-dependent Kondo coupling $J_{\bold k,\bold k',i}=\frac{1}{N}\frac{V_{\bold k}^{*} V_{\bold k'}e^{i(\bold k-\bold k')\cdot R_i}}{\epsilon_{p\bold k}+\Delta_s}$, which contains the information of the hybridization between $s$ and $p_X$ orbitals.

\begin{figure}[t]
\centering
\includegraphics[width=3in]{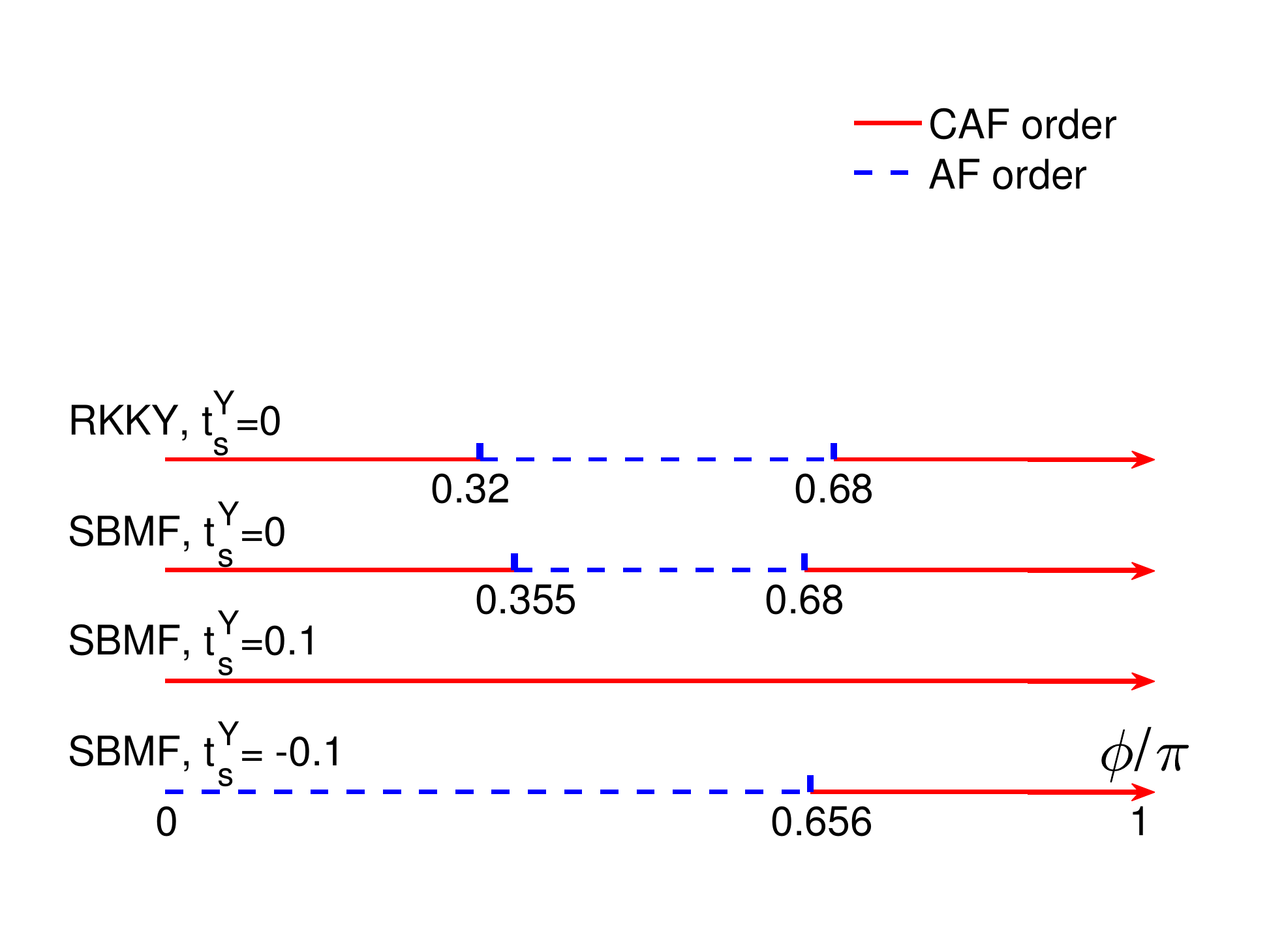}
\caption{One-dimensional magnetic phase diagrams with $(t_p^X,t_{sp},\Delta _s)=(1,0.3,3)$ determined by phase $\phi$. The phase diagram on the top is obtained from RKKY interaction by perturbation theory with $t_s^Y=0$, and three lower phase diagrams are obtained with slave-boson mean-field theory (SBMF) with different $t_s^Y$. The red solid lines correspond to CAF order with $Q=(\pi,0)$ and blue dashed lines correspond to AF order with $Q=(\pi,\pi)$. }
\label{tsp0d3phase}
 \end{figure}

\subsection{The RKKY interaction and static spin susceptibility}
The RKKY interaction is derived from the Kondo lattice model in weak Kondo coupling $J$ regime by second-order perturbation at the cost of eliminating the itinerant electron degree of freedom. Although the Kondo effect is omitted in such perturbation treatment, the RKKY interaction is helpful for searching possible magnetic orders. We now derive the RKKY interaction based on the Kondo lattice model obtained in the last subsection by applying the projection operator method again. In the perturbation treatment, it is assumed that $p_X$-band is treated as free band without coupling to the $s$-orbitals, while the $s$-$p_X$ hybridization shall be considered up to second order to derive RKKY interaction for $s$-orbitals. Thus the $p_X$-orbital has the own Fermi level $\epsilon_{p, k_F}$. The $p_X$-orbital states have fermion distribution $n_{p,k}=1$ if $\epsilon_{p,k}<\epsilon_{p,k_F}$, and $n_{p,k}=0$ if $\epsilon_{p,k}>\epsilon_{p,k_F}$. We further define that the projection operator $P_0$ projects states onto the subspace with a ground state Fermi sea formed by the $p_X$-orbital degree of freedom, so in subspace $P_0$ no particle excitation above Fermi sea or hole excitation below Fermi sea exists. By the second order perturbation (see details in Appendix) similar to that in the above subsection, we obtain the RKKY interaction:
\begin{widetext}
   \begin{equation}
    \begin{aligned}
H_{RKKY}= \sum\limits_{i,j}2J(X_i-X_j,Y_i-Y_j)\bold{S}_i\cdot \bold{S}_j,
   \end{aligned}
   \label{rkkyh}
  \end{equation}
where the coupling coefficient takes the form
     \begin{equation}
    \begin{aligned}
J(X_i-X_j,Y_i-Y_j)=\sum\limits_{\bold k,\bold k'}\frac{4\cos[(\bold k-\bold k')\cdot(\bold R_i-\bold R_j)]}{N^2}|V_{\bold k}|^2|V_{\bold k'}|^2\frac{1}{\epsilon_{p\bold k}+\Delta_s}\frac{1}{\epsilon_{p\bold k'}+\Delta_s}\frac{n_{p,\bold k}-n_{p,\bold k'}}{\epsilon_{p\bold k}-\epsilon_{p\bold k'}}.
   \end{aligned}
  \end{equation}

The static spin susceptibility $\chi (\bold Q)$, which is the Fourier transformation of the real space coupling coefficient $J(X_i-X_j,Y_i-Y_j)$, can then be obtained directly:
\begin{equation}
    \begin{aligned}
\chi (\bold Q)
=\sum\limits_{\bold k}2|V_{\bold k}|^2|V_{\bold k+\bold Q}|^2\frac{1}{\epsilon_{p,\bold k}+\Delta_s}\frac{1}{\epsilon_{p,\bold k+\bold Q}+\Delta_s}\frac{n_{p,\bold k}-n_{p,{\bold k}+\bold Q}}{\epsilon_{p,\bold k}-\epsilon_{p,\bold k+\bold Q}}.
   \end{aligned}
   \label{suscepti}
  \end{equation}
  \end{widetext}

The $\chi (\bold Q)$ is just the dispersion relation of the RKKY interaction Hamiltonian~\cite{wxg} if we view the quantum spin model Eq.~\eqref{rkkyh} as a classical spin model, and the vector $Q$ which minimizes $\chi (\bold Q)$ is the ground state magnetic order of the corresponding classical spin model.

From Eq.~\eqref{suscepti} one can find the susceptibility function of $Q$ is different from that in a standard Kondo lattice model. On one hand, the hybridization function $|V_{\bold k}|^2=4t_{sp}^2[\cos\phi (\cos(k_X)+\cos(k_Y))+\cos(k_X)\cos(k_Y)+1]$ is $\bold k$ and $\phi$-dependent, originating from the feature of the superlattice that each $p_X$ orbital resides in the center of four nearest neighbour $s$ orbitals and can hop directly to one of them with phase. On the other hand, the dispersion of the itinerant $p_X$ band is highly anisotropic and relevant only in one-dimensional, while the model is two-dimensional due to the two-dimensional hybridization. For simplicity we replace the term $1/(\epsilon_{p,\bold k}+\Delta_s)$ with $1/(\epsilon_{p,\bold k}+\Delta_s)$, since the particle scattering mostly occurs near the Fermi level $\epsilon_{p,\bold k_F}$. We can then find the RKKY interaction with $\phi$ is equivalent to that with $\pi-\phi$ and the magnetic phase diagram from this approach is symmetric about $\phi=\pi/2$.

To see clearly the magnetic effects from the susceptibility Eq.~\eqref{suscepti}, we separately look at the contributions from the Fermi surface nesting term
 \begin{equation}
 \begin{aligned}
\sum\limits_{\bold k}\frac{n_{p,\bold k}-n_{p,{\bold k}+\bold Q}}{\epsilon_{p,\bold k}-\epsilon_{p,\bold k+\bold Q}},
 \end{aligned}
 \label{nesting}
 \end{equation}
  and from the hybridization term
 \begin{equation}
 \begin{aligned}
&\sum\limits_{\bold k}|V_{\bold k}|^2|V_{\bold k+\bold Q}|^2\\
\propto&- \cos{Q_X}\cos{Q_Y}-2\cos{\phi}^2( \cos{Q_X}+\cos{Q_Y})-4.
\end{aligned}
 \label{vq}
 \end{equation}

\begin{figure}[t]
\centering
\includegraphics[width=3in]{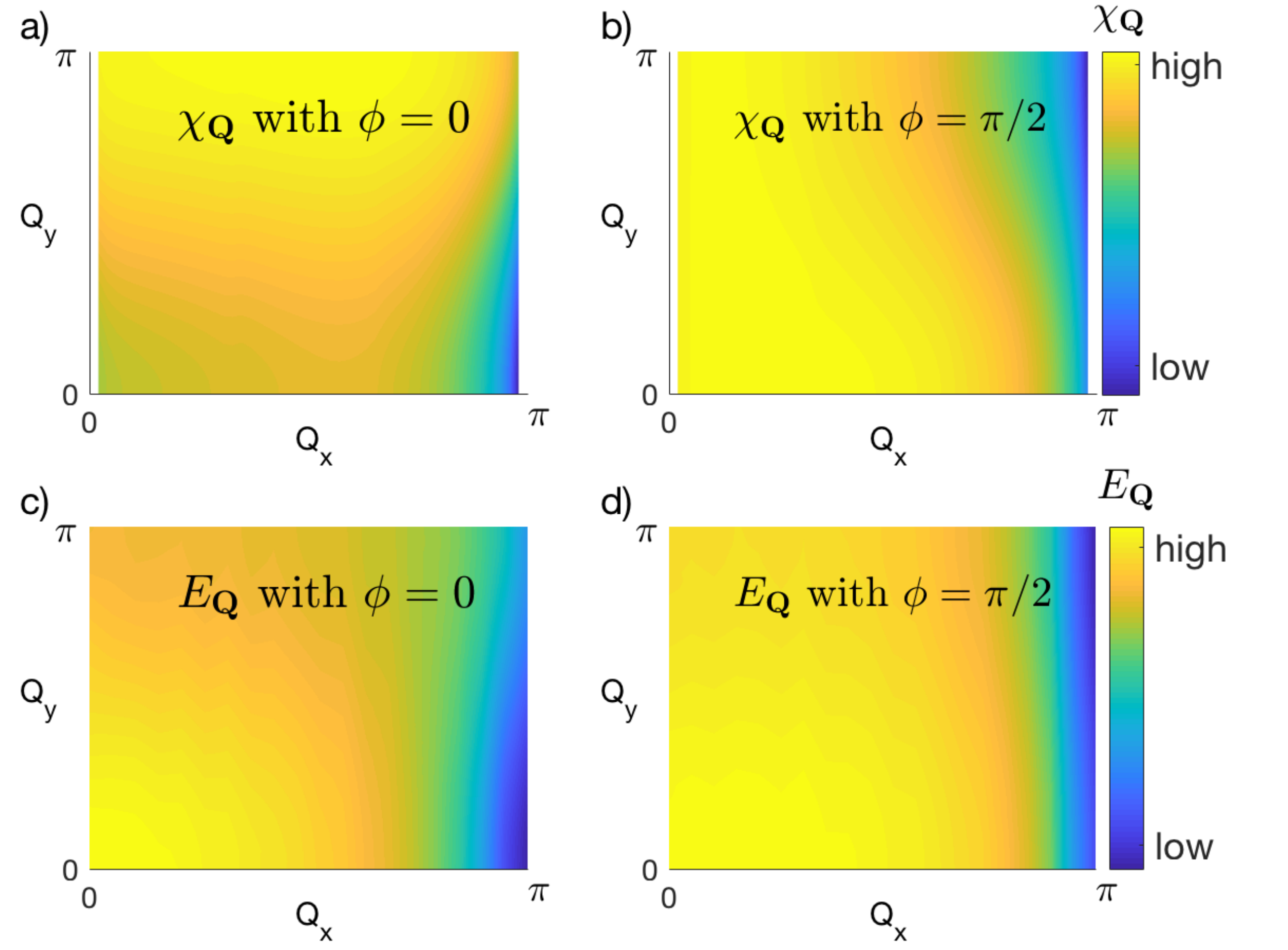}
\caption{Magnetic susceptibility/slave-boson mean-field energy as functions of magnetic order $\bold Q$ with $(t_s^Y,t_p^X,t_{sp},\Delta _s)=(0,1,0.5,3)$ for $\phi=0$ and $\phi=\pi/2$. The dark blue and yellow colors represent the minimum and maximum of the susceptibility/energy, respectively. a)-b) Static susceptibility $\chi_{\bold Q}$ obtained from RKKY interaction by perturbation theory for $\phi=0$ and $\phi=\pi/2$. In the limit $Q_X=\pi$ the susceptibility diverges. c)-d) Mean-field energy $E_{\bold Q}$ obtained with slave-boson mean-field theory for $\phi=0$ and $\phi=\pi/2$. For a) and c), the ground state magnetic order is $(\pi,0)$; for b) and d), the ground state magnetic order is $(\pi,\pi)$.}
\label{susceptibility}
 \end{figure}

It can be seen that the Fermi surface nesting term Eq.~\eqref{nesting} tends to result in antiferromagnetic order (AF) magnetic order with $Q_X=2k_F=\pi$ in $\hat{X}$ direction since the $p_X$ orbitals only hop in $\hat{X}$ direction and the band formed by $p_X$ orbitals is half filled. On the other hand, the effect of the hybridization term Eq.~\eqref{vq} depends on the phase $\phi$. For $\phi=\pi/2$, Eq.~\eqref{vq} equals to $-\cos{Q_X}\cos{Q_Y}-4$ and favors the magnetic order with $\bold Q=(0,0)$ or $(\pi,\pi)$; while for $\phi=0$ or $\pi$, Eq.~\eqref{vq} equals to $-(\cos{Q_X}+2)(\cos{Q_Y}+2)$ and favors the magnetic order the $\bold Q=(0,0)$. The order $\bold Q=(0,0)$ therefore competes with the Fermi surface nesting effect in $\hat{X}$ direction. We numerically calculated Eq.~\eqref{suscepti} and plotted the one-dimensional magnetic phase diagram in [Fig.~\ref{tsp0d3phase}]. The figure shows that near $\phi=\pi/2$ the order is AF with $\bold Q=(\pi,\pi)$ while near $\phi=0$ or $\pi$ the order is collinear antiferromagnetic order (CAF) with $\bold Q=(\pi,0)$, implying that the Fermi surface nesting effect dominates in weak hybridization regime. We also plotted a special case of Eq.~\eqref{suscepti} with $(t_s^Y,t_p^X,t_{sp},\Delta _s)=(0.1,1,0.5,3)$ as a function of $Q_X$ and $Q_Y$ in [Fig.~\ref{susceptibility} a)-b)]. Note that at $Q_X=\pi$ in Eq.~\eqref{nesting} and the susceptibility diverges due to the one-dimensional character of $p_X$ band dispersion, although the magnetic order is two-dimensional due to the two-dimensional hybridization.

\begin{figure}[t]
\centering
\includegraphics[width=3in]{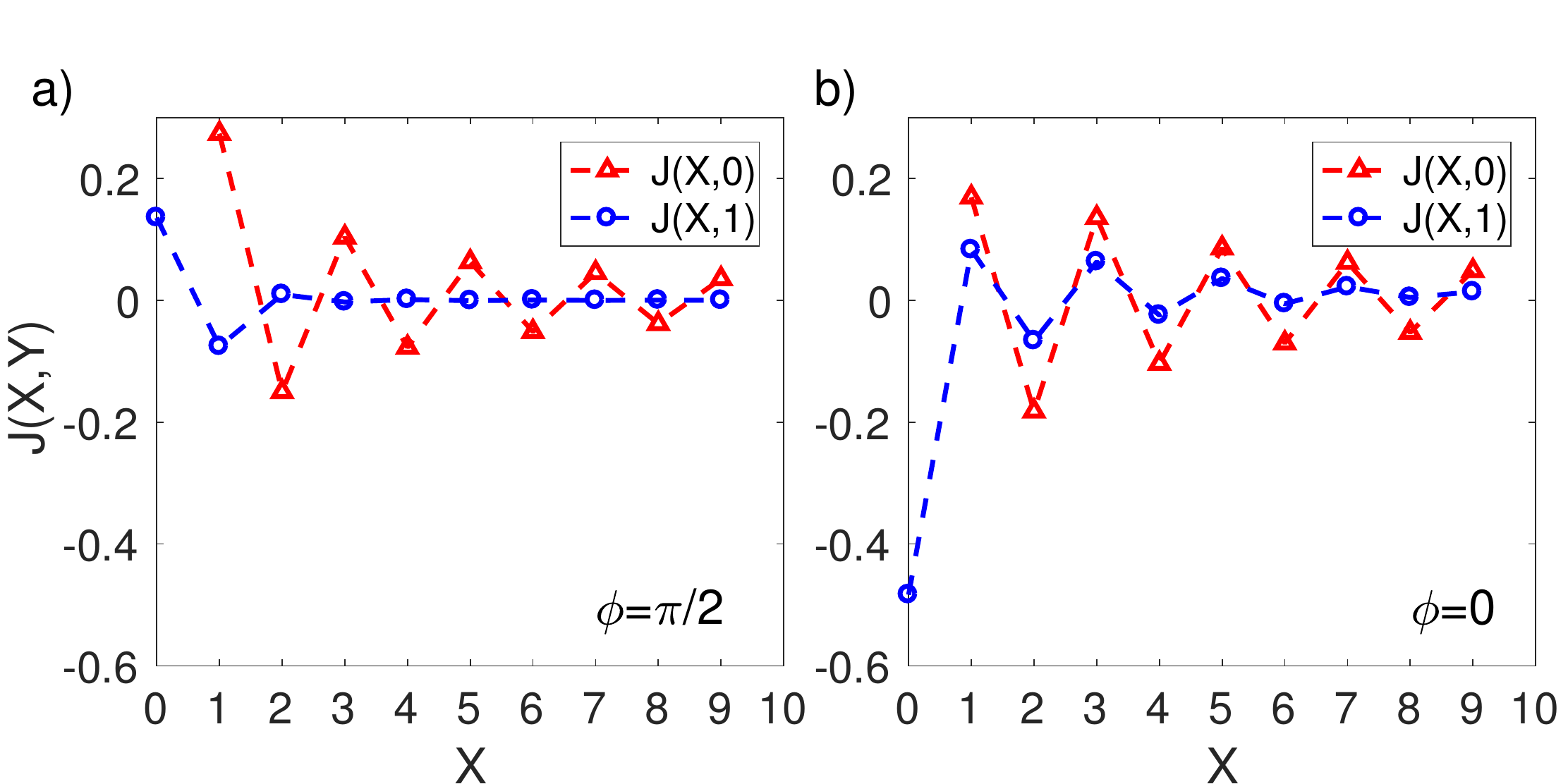}
\caption{The RKKY coupling coefficient $J(X,Y)=J(X_i-X_j,Y_i-Y_j)$ with $(t_s^Y,t_p^X,\Delta _s)=(0,1,3)$ for a) $\phi=\pi/2$ and b) $\phi=0$ in the unit of $|J(0,0)|$.}
\label{Jxy}
 \end{figure}

We further investigate the properties of RKKY interaction in real space from Eq.~\eqref{rkkyh}. In $\hat{Y}$ direction, coupling coefficients with $|Y_i-Y_j|>1$ always vanish due to the anisotropic $p_X$ band and the RKKY magnetic interaction is the 4-th order virtual process with respect to $t_{sp}$. While in $\hat{X}$ direction, as the value of Eq.~\eqref{nesting} diverges at $Q_X=\pi$ due to one-dimensional character of $\epsilon_{p\bold k}$, the coupling coefficients decays slowly. As a result, the coupling coefficients are short-ranged in $\hat{Y}$ direction and long-ranged in $\hat{X}$ direction. We numerically calculated the coupling coefficients $J(X_i-X_j,Y_i-Y_j)$ with Eq.~\eqref{rkkyh} for $\phi =\pi/2$ and $0$ as shown in [Fig.~\ref{Jxy}]. The signs of $J(X,0)$ in the two cases are the same and favors $Q_X=\pi$ order. However, the signs of $J(X,1)$ in the two cases differ by $-1$, and it can be seen that $Q_Y=\pi$ is supported when $\phi =\pi/2$ while $Q_Y=0$ is supported when $\phi =0$, which is consistent with [Fig.~\ref{tsp0d3phase}]. As the signs of $J(X,Y)$ oscillate, there is no geometric frustration in these two special cases. From these results, we obtain that when $\phi$ is tuned from $0$ to $\pi/2$ or from $\pi$ to $\pi/2$, the magnetism has a transition from CAF order to AF order, as shown in Fig.~\ref{tsp0d3phase}.

\section{Magnetic phases and their effects on QAH Kondo states}

Note that the RKKY interaction, which shows the possible magnetic orders, is obtained by perturbation theory and valid in weak hybridization regime. Moreover, the derivation of the effective RKKY interaction is at the cost of eliminating the $p_X$ orbital degree of freedom, which fails to study Kondo and QAH effect since only $s$-orbital is left. To overcome these drawbacks, in this section we apply the non-perturbative slave-boson mean-field theory to study the ground state phases and magnetic effects on the QAH effect in our CK model. To take into account the magnetic orders suggested by the RKKY interaction, we apply the spin-rotation invariant slave-boson mean-field theory~\cite{rotation} in the subsection A, which is convenient to describe various magnetic orders. In subsection B and C, we show the magnetic and correlated QAH phase diagrams in Fig.~\ref{phase}, and in subsection D we discuss how to identify the influences of strong correlation and magnetism on the CK phase by measuring Hall conductance and double occupancy in cold atom experiments. Different from the previous work~\cite{chernkondo}, we show the rich magnetic phases coexisting with Kondo hybridization and find that in the magnetic Kondo phases the effective hybridization, and then the correlated QAH phase, are enhanced compared with the paramagnetic Kondo phase in relatively weak hybridization regime.

\subsection{Slave-boson mean-field theory}

In the present CK model (periodic Anderson model), due to the strong repulsive Hubbard interaction in $s$ orbitals, the double occupancy of $s$ orbitals is suppressed to zero. Such a system can be studied with slave-boson theory~\cite{colemansb} proposed by Coleman. Kotliar and Ruckenstein (KR) futher extended the Coleman slave-boson representation to a more complex form~\cite{sb} that incorporates the result of the Gutzwiller approximation~\cite{gutzwiller} on the mean-field level. In this paper we apply the spin-rotation invariant slave-boson mean-field theory~\cite{rotation} which is a generalized form of the KR slave-boson theory and is convenient to explore various magnetic orders.

In the spin-rotation invariant slave-boson theory~\cite{rotation}, the auxiliary bosonic and fermionic operators can be introduced. For this we introduce the slave-boson operators $\hat{e},\hat{d},\hat{p}_0, \hat{\bold p}=(\hat{p}_1,\hat{p}_2,\hat{p}_3)$ that obey bosonic commutation relation. Here $\hat{e}, \hat{d}$ correspond to hole and doubly occupied states. The scalar ($S=0$) field $\hat{p}_0$ and vector ($S=1$) field $\hat{\bold p}=(\hat{p}_1,\hat{p}_2,\hat{p}_3)$ correspond to the singly occupied state. Note that $\hat{e},\hat{d},\hat{p}_0$ transform as scalars under spin rotation, while $\hat{\bold p}$ transforms as a vector. On the other hand, the $S =1/2$ pseudo-fermion operators $c_{i\sigma }$ obey fermionic commutation relation. The key idea is that the singley occupied auxiliary bosonic and fermionic modes shall form into spin-$1/2$ $s$-orbtial fermion states under proper constrains. The local $s$ orbital operators $s_{\sigma}$ are then represented by $s_{\sigma}=\sum_{\sigma '}\hat{z}_{\sigma\sigma '}c_{\sigma '}$, with the matrix $\underline{z}$ defined as (see more details in Appendix)
 \begin{equation}
  \begin{aligned}
\underline{\hat{z}}=\hat{e}^{\dagger}\underline{L} \underline{R}\underline{\hat{p}}+\underline{\hat{\tilde{p}}}^{\dagger}\underline{L}\underline{R}\hat{d},
    \end{aligned}
  \end{equation}
  where
\begin{equation}
  \begin{aligned}
&\underline{L}=\left[ (1-\hat{d}^{\dagger}\hat{d})\underline{1}-2\underline{\hat{p}}^{\dagger}\underline{\hat{p}}\right]^{-\frac{1}{2}},\\
&\underline{R}=\left[ (1-\hat{e}^{\dagger}\hat{e})\underline{1}-2\underline{\hat{\tilde {p}}}^{\dagger}\underline{\hat{\tilde{p}}}\right]^{-\frac{1}{2}}.\\
    \end{aligned}
  \end{equation}
Here $\underline{\hat{z}},\underline{L},\underline{R}$ are $2\times 2$ matrices, matrix elements of the matrix $\underline{\hat{p}}$ is defined as $\hat{p}_{\sigma \sigma '}=\frac{1}{2}\sum_{\mu=0}^3 \hat{p}_{\mu}\tau_{\mu,\sigma\sigma '}$, and its time reversal transformation reads $\hat{\tilde{p}}_{\sigma \sigma '}=(\hat{T}\underline{\hat{p}}\hat{T}^{-1})_{\sigma \sigma '}=\sigma \sigma ' \hat{p}_{\bar{\sigma}'\bar{\sigma}}$. For each $s$ orbital at site $\bold R_i$, a set of above auxiliary operators are induced with index $i$ labeling their sites. The total Hilbert space has been extended now and the physical subspace can be obtained through following constraints:
 \begin{equation}
  \begin{aligned}
 & \hat{e}_i^{\dagger}\hat{e}_i+\hat{d}_i^{\dagger}\hat{d}_i+\sum\limits_{\mu}\hat{p}_{i\mu}^{\dagger}\hat{p}_{i\mu}-1=0,\\
 & \sum\limits_{\sigma}c_{i\sigma}^{\dagger}c_{i\sigma}-\sum\limits_{\mu} \hat{p}_{i\mu}^{\dagger} \hat{p}_{i\mu}-2\hat{d}_i^{\dagger }\hat{d}_i=0,\\
&\sum\limits_{\sigma\sigma'}\boldsymbol \tau_{\sigma\sigma'} c_{i\sigma '}^{\dagger}c_{i\sigma}-\hat{p}_{i0} \hat{\boldsymbol p}_i^{\dagger}-\hat{\boldsymbol  p}_i^{\dagger}\hat{p}_{i0}
+i (\hat{\bold p}_i^{\dagger} \times \hat{\bold p}_i)=0.
    \end{aligned}
  \label{constraints}
  \end{equation}

In terms of the auxiliary operators and incorporating the constraints in form of Lagrange multiplier fields $\alpha_i, {\beta}_{i0}$ and ${\boldsymbol \beta}_i$, the CK Hamiltonian takes the form
\begin{widetext}
\begin{equation}
 \begin{aligned}
H=&\sum\limits_{i\sigma} \left[ \sum\limits_{\sigma' \sigma ''}t_{s}^Y \hat{z}_{i\sigma \sigma '}^{\dagger}\hat{z}_{i\pm\hat{Y}\sigma '' \sigma} c_{i\sigma'}^{\dagger} c_{i\pm\hat{Y}\sigma''}-\Delta_s    c_{i\sigma}^{\dagger} c_{i\sigma}+ t_{p}^Xp_{Xi\sigma}^{\dagger} p_{Xi\pm\hat{X}\sigma}\right]+\left[\sum_{\langle ij\rangle \sigma}F(\bold r) \hat{z}_{i\sigma \sigma '}^{\dagger}c_{i\sigma '}^{\dagger}p_{Xj\sigma}\delta_{j,i+\bold r}+\mathrm{H.c.}\right ]\\
&+\sum\limits_{i}    \bigg[ U_s \hat{d}_i^{\dagger}\hat{d}_i +\alpha_{i}(\hat{e}_i^{\dagger}\hat{e}_i+\hat{d}_i^{\dagger}\hat{d}_i+\sum\limits_{\mu}\hat{p}_{i\mu}^{\dagger}\hat{p}_{i\mu}-1)
+{ \beta}_{i0} (\sum\limits_{\sigma}c_{i\sigma}^{\dagger}c_{i\sigma}-\sum\limits_{\mu} \hat{p}_{i\mu}^{\dagger} \hat{p}_{i\mu}-2\hat{d}_i^{\dagger }\hat{d}_i)\\
&+{\boldsymbol \beta}_i\cdot (\sum\limits_{\sigma\sigma'}\boldsymbol \tau_{\sigma\sigma'} c_{i\sigma '}^{\dagger}c_{i\sigma}-\hat{p}_{i0} \hat{\boldsymbol p}_i^{\dagger}-\hat{\boldsymbol  p}_i^{\dagger}\hat{p}_{i0}
+i (\hat{\bold p}_i^{\dagger} \times \hat{\bold p}_i)) \bigg].
  \end{aligned}
  \label{Hamiltonian}
  \end{equation}
\end{widetext}

We consider the mean-field approximation to the boson fields in the above Hamiltonian with infinitely large $s$ orbital on-site interaction $U_s$. In this case, the scalar bosonic mean-field order vanishes $ d_i=0$, the scalar mean-field orders $e_i, p_{0i}$ can be assumed spatially uniform so that $e_i=e$ and $p_{0i}=p_0$, and also the Lagrange multiplier fields $\alpha_i=\alpha, \beta_{0i}=\beta_0$. We further consider that the vector mean-field orders take the forms $\bold p_i=p(\cos{(\bold Q\cdot \bold R_i)},\sin{(\bold Q\cdot \bold R_i)},0)$ and $\boldsymbol \beta_i=\beta(\cos{(\bold Q\cdot \bold R_i)},\sin{(\bold Q\cdot \bold R_i)},0)$, characterizing a spatially rotation structure in $\hat{X}-\hat{Y}$ plane. The magnetic order wave-vector $\bold Q$ can be commensurate or incommensurate. In particular, $\bold Q=(\pi,\pi)$ denotes AF order, while the order is incommensurate if any component of $\bold Q/\pi$ is irrational. Based on the above assumptions, the mean-field Hamiltonian (see details in Appendix) in the basis $\bold X_{\bold k}^{\dagger}\equiv (c_{\bold k\uparrow}^{\dagger},c_{\bold k+\bold Q\downarrow}^{\dagger},p_{X\bold k\uparrow}^{\dagger},p_{X\bold k+\bold Q\downarrow}^{\dagger})$ takes the form
\begin{equation}
  \begin{aligned}
  H=&\sum\limits_k \bold X_{\bold k}^{\dagger} \epsilon_k \bold X_{\bold k}+N \Big[-\beta_0(p_0^2+p^2)\\
  &+2\beta p_0p+\alpha(e^2+p^2+p_0^2-1)         \Big ],
\end{aligned}
 \end{equation}
with matrix $\epsilon_k$ defined as:
\begin{equation}
\epsilon_{k}  =
  \left(
  \begin{matrix}
  \epsilon_{s\bold k}^a+\beta_0 &\epsilon_{s\bold k }^c+\beta & z_+V_{\bold k}& z_-V_{\bold k+\bold Q}\\
       \epsilon_{s\bold k }^c+\beta  &\epsilon_{s\bold k}^b+\beta_0& z_-V_{\bold k}&z_+V_{\bold k+\bold Q}\\
        z_+V_{\bold k}^*& z_-V_{\bold k}^*&\epsilon_{p,\bold k}&0\\
      z_-V_{\bold k+\bold Q}^* &z_+V_{\bold k+\bold Q}^*&0&\epsilon_{p,\bold k+\bold Q}
  \end{matrix}
  \right).
  \label{matrix}
  \end{equation}
Here $\epsilon_{s\bold k}^a=z_+^2\epsilon_{s\bold k}+z_-^2\epsilon_{s\bold k+\bold Q}-\Delta_s$, $\epsilon_{s\bold k}^b=z_+^2\epsilon_{s\bold k+\bold Q}+z_-^2\epsilon_{s\bold k}-\Delta_s$, $\epsilon_{s\bold k}^c=z_+z_-(\epsilon_{s\bold k+\bold Q}+\epsilon_{s\bold k})$ are $s$ orbital hopping terms. The renormalization factor takes the form~\cite{hubbard}
\begin{equation}
  \begin{aligned}
z_{\pm}=&\frac{1}{\sqrt{2}}\frac{ep_{+}}{\sqrt{1-p_+^2}\sqrt{1-e^2-p_-^2}}\\
&\pm \frac{1}{\sqrt{2}}\frac{ep_-}{\sqrt{1-p_-^2}\sqrt{1-e^2-p_+^2}},
 \end{aligned}
 \label{zplus}
 \end{equation}
  where $p_{\pm}=(p_0\pm p)/\sqrt{2}$ are proportional to the eigenvalues of $\hat{p}_{\sigma\sigma'}$ matrix.

The mean-field solutions (saddle point approximation) are obtained by minimizing the mean-field free energy (ground state energy at zero temperature), which reads
 \begin{equation}
  \begin{aligned}
  F=&-T\sum\limits_{\bold k\alpha} \ln\big [1+\exp(-(E_{\bold k \alpha}-\mu)/T)\big ]+\\
  &N \Big[-\beta_0(p_0^2+p^2)
  +2\beta p_0p+\alpha(e^2+p^2+p_0^2-1) \Big ].
   \end{aligned}
  \end{equation}
Here the $\epsilon_{\bold k \alpha}$ with $\alpha=1,2,3,4$ are four eigenvalues of the $4\times 4$ matrix $\epsilon_{\bold k}$. The stationary condition yields the saddle point equations
  \begin{equation}
  \begin{aligned}
  \frac{\partial F}{\partial e} = \frac{\partial F}{\partial p_0}= \frac{\partial F}{\partial p}= \frac{\partial F}{\partial \alpha}= \frac{\partial F}{\partial \beta_0}= \frac{\partial F}{\partial \beta}=0.
  \end{aligned}
  \end{equation}
The chemical potential $\mu$ at half filling is determined by the particle number condition
 \begin{equation}
  \begin{aligned}
\sum\limits_{\bold k \alpha}f(E_{\bold k \alpha}-\mu)=2N,
 \end{aligned}
 \end{equation}
where $f$ is the Fermi-Dirac distribution function function. Since we only consider ground state in this paper, the distribution function is just a step function.

The physical quantities such as particle numbers and magnetic moments can be represented by correlation functions. Particle number and magnetic moment in $s$ orbitals are obtained by
 \begin{equation}
  \begin{aligned}
&n_{s\uparrow}=\sum\limits_{\bold k \in BZ}\langle c_{\bold k\uparrow}^{\dagger}c_{\bold k\uparrow}\rangle,\\
&n_{s\downarrow}=\sum\limits_{\bold k \in BZ}\langle c_{\bold k\downarrow}^{\dagger}c_{\bold k\downarrow}\rangle,\\
&\bold m_{s}=\sum\limits_{\bold k \in BZ,\sigma\sigma '}\langle c_{\bold k\sigma}^{\dagger}\boldsymbol \tau_{ \sigma \sigma'}c_{\bold k+\bold Q\sigma}\rangle.\\
 \end{aligned}
 \end{equation}

Similarly, the particle number and magnetic moment in $p_X$ orbitals take the form
 \begin{equation}
   \begin{aligned}
&n_{p\uparrow}=\sum\limits_{\bold k \in BZ}\langle p_{X\bold k\uparrow}^{\dagger}p_{X\bold k\uparrow}\rangle,\\
&n_{p\downarrow}=\sum\limits_{\bold k \in BZ}\langle p_{X\bold k\downarrow}^{\dagger}p_{X\bold k\downarrow}\rangle,\\
&\bold m_{p}=\sum\limits_{\bold k \in BZ,\sigma\sigma '}\langle p_{X\bold k\sigma}^{\dagger}\boldsymbol\tau_{ \sigma \sigma'}p_{X\bold k+\bold Q\sigma}\rangle.\\
 \end{aligned}
 \end{equation}

 The first Chern number $ \text{Ch}_1$ for the bands below Fermi level $\mu$ is calculated by
\begin{equation}
 \begin{aligned}
 \text{Ch}_1 &=\int_{BZ}d^2\bold k\sum\limits_{\alpha, \alpha'}[f(E_{\bold k \alpha'}-\mu)-f(E_{\bold k \alpha}-\mu)]\\
 &\times \frac{1}{2\pi}  \frac{\text{Im}[\langle \bold k \alpha '|\hat{v}_X| \bold k \alpha \rangle   \langle \bold k \alpha |\hat{v}_Y| \bold k \alpha ' \rangle  }{(E_{\bold k \alpha '}-E_{\bold k \alpha})^2},
\end{aligned}
\end{equation}
where $\hat{v}_{X/Y}=\partial \epsilon_{\bold k}/\partial k_{X/Y}$ are velocity operators along $\hat{X}/\hat{Y}$ directions and $|\bold k \alpha\rangle$ is the eigenvector of matrix $\epsilon_{\bold k}$ corresponding to the eigenenergy $E_{\bold k \alpha}$.

\subsection{Magnetic phase diagrams}

The mean-field saddle point equations can be solved numerically with an ansatz magnetic order $\bold Q=(Q_X,Q_Y)$, and the saddle point solution yields the mean-field ground state energy $E_{\bold Q}$. The ground state magnetic order should be obtained by finding $\bold Q$ that minimizes the mean-field energy $E_{\bold Q}$.

As suggested by the features of the RKKY interaction discussed in the last section, the $s$-$p_X$ hybridization favors the order $\bold Q=(0,0)$ or $(\pi,\pi)$ and the Fermi surface nesting effect in $p_X$ band favors the order $Q_X=\pi$. As a result, we consider only the paramagnetic, ferromagnetic ($\bold Q=(0,0)$), AF ($\bold Q=(\pi,\pi)$), and CAF ($Q=(\pi,0)$) orders in the slave-boson mean-field calculation. In Fig.~\ref{susceptibility} c)-d) we plot $E_{\bold Q}$ versus $\bold Q$ for the parameters $(t_{sp},t_s^Y,t_p^X,t_{sp},\Delta _s)=(0.5,0,1,0.5,3)$ and $\phi=0$, $\pi/2$ respectively. One can see that the relation between $E_{\bold Q}$ and $\bold Q$ in low energy regions shown in Fig.~\ref{susceptibility} c)-d) qualitatively agrees with the relation between the static magnetic susceptibility $\chi_{\bold Q}$ and $\bold Q$ in Fig.~\ref{susceptibility} a)-b), and the ground state magnetic orders obtained from both methods are the same. However, in the strong coupling $t_{sp}$ regime, the magnetic orders from perturbation theory may not be justified, since the perturbation theory is not valid in strong hybridization regime.

 \begin{figure*}[t]
\centering
\includegraphics[width=17cm, height=6cm]{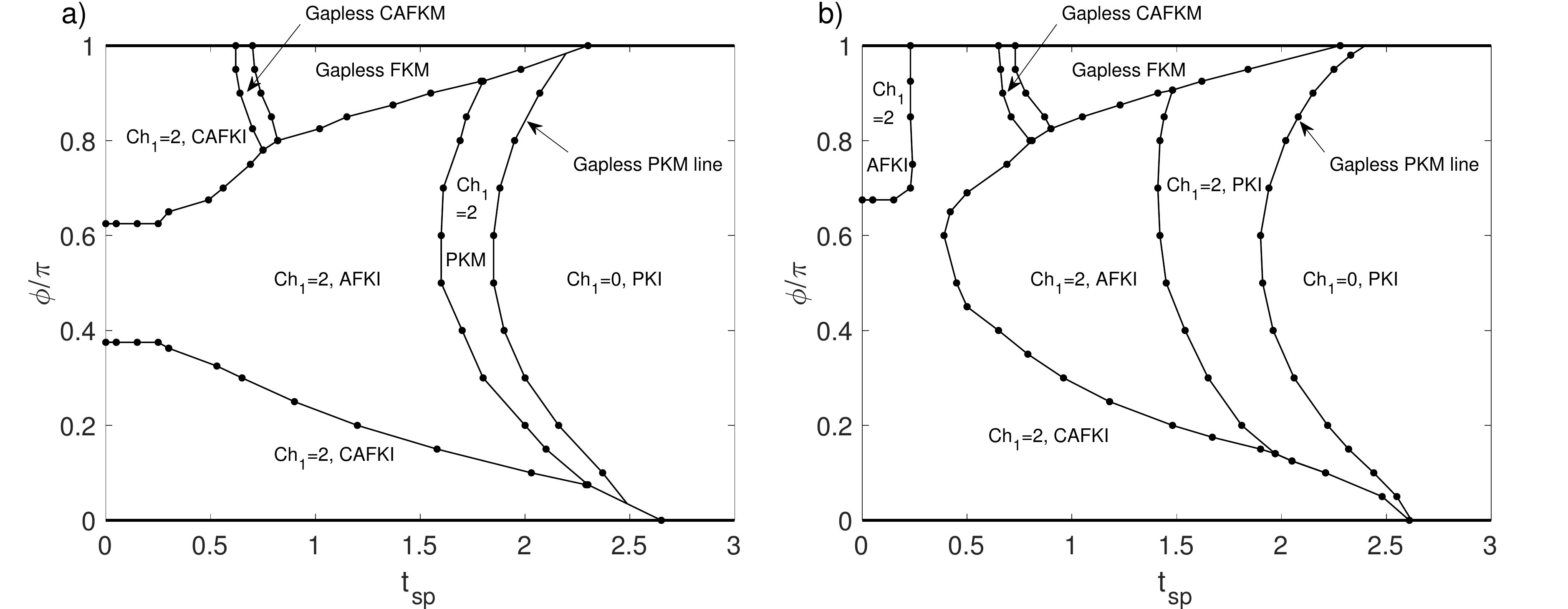}
\caption{The magnetic and QAH phase diagrams based on slave-boson theory for $(t_p^X,\Delta _s)=(1,3)$ for $t_s^Y=0$ in a) and $t_s^Y=0.1$ in b), respectively. The $\text{Ch}_1$ refers to as the first Chern number. The phases shown here include antiferromagnetic Kondo insulator (AFKI), collinear antiferromagnetic Kondo insulator (CAFKI), collinear antiferromagnetic Kondo metal (CAFKM), ferromagnetic Kondo metal (FKM), paramagnetic Kondo insulator (PKI), and paramagnetic Kondo metal (PKM) phases. The dotted lines represent phase boundaries. The Chern number for $\phi=0$ and $\phi=\pi$ lines are always zero but is not displayed in the figure.}
\label{phase}
 \end{figure*}

\begin{figure}[t]
\centering
\includegraphics[width=3in]{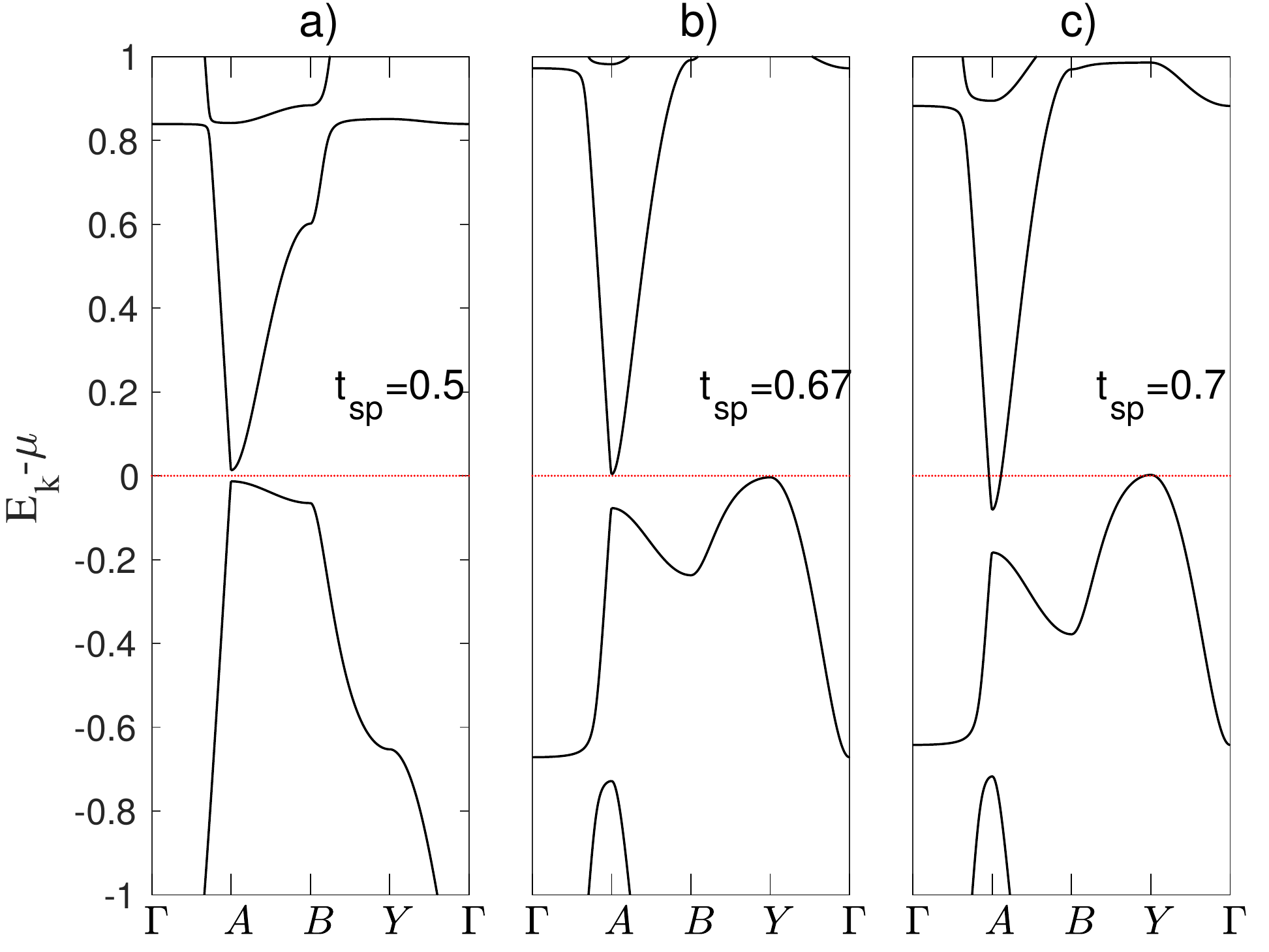}
\caption{The quasiparticle band obtained with slave-boson theory shows a transition from CAFKI phase to CAFKM phase at critical value $t^c_{sp}\simeq 0.67$ for $(t_s^Y,t_p^X,\Delta_s,\phi)=(0.1,1,3,0.9\pi)$. The special $k$-points are defined as $\Gamma=(0,0),A=(\pi/2,0),B=(\pi/2,\pi),Y=(0,\pi)$.  }
\label{band}
 \end{figure}

We plot the ground state magnetic phase diagrams versus the coupling $t_{sp}$ and phase $\phi$ in Fig.~\ref{phase} with $(t_p^X,\Delta _s)=(1,3)$. Note that the phase diagram from the present mean-field calculation is symmetric about $\phi=\pi$, so for convenience we plot only the phase diagrams in the region $0<\phi<\pi$. Fig.~\ref{phase} a) shows the phase diagram with $t_s^Y=0$, where the Hamiltonian can be transformed to the Kondo lattice Hamiltonian in weak hybridization regime as discussed previously. In weak hybridization regime $t_{sp}\lesssim 0.6$ the magnetic phase diagram qualitatively agrees with the perturbation result in Fig.~\ref{tsp0d3phase}. Namely, near $\phi=\pi/2$ the phase is AF Kondo insulator (AFKI), while near $\phi=0$ and $\phi=\pi$ the phase is CAF Kondo insulator (CAFKI). At large coupling regime, the slave-boson theory shows that for $\phi\approx \pi$, the phase evolves from CAFKI phase to the CAF Kondo metal (CAFKM) phase as $t_{sp}$ increases (band structures is shown in Fig.~\ref{band}), and then to ferromagnetic Kondo metal (FKM) phase. We note that the ferromagnetic order is not a ground-state order at half filling in standard periodic Anderson model (with simple isotropic conduction band dispersion and $k$ independent hybridization) and may be ground state only away from half filling in slave-boson mean-field calculation~\cite{sbmf}. However, in the present CK model the emergence of ferromagnetism at half filling is due to the anisotropic $k$-dependent hybridization function $V_{\bold k}$. In the region $\phi\approx 0$, the increase of $t_{sp}$ does not affect the phase diagram much and the magnetic order is always CAF before magnetic moment decreases to zero. This asymmetry of the magnetic phase diagram about $\phi=\pi/2$ is in contrast to the symmetry of the magnetic phase diagram from RKKY interaction in last section, where the second order perturbation eliminates the $p_X$ orbitals' degree of freedom. At very strong hybridization, the magnetic Kondo phases evolve into paramagnetic Kondo metal (PKM) phase, and further into paramagnetic Kondo insulator (PKI) phase, consistent with the Doniach diagram that the magnetic orders are fully suppressed in the strong coupling regime~\cite{doniach}.

Fig.~\ref{phase} b) shows the magnetic phase diagram with $t_s^Y=0.1$. The main difference in this case from that with $t_s^Y=0$ [as shown in Fig.~\ref{phase} a)] lies in weak hybridization $t_{sp}$ regime. With $t_s^Y=0.1$, the AF order disappears when $t_{sp}\lesssim 0.4$ but again appears at very weak $t_{sp}$ and $\phi\approx\pi$. A comparison of magnetic orders with parameters $(t_p^X,t_{sp},\Delta _s)=(1,0.3,3)$ for $t_s^Y=0$ and $\pm 0.1$ is shown in [Fig.~\ref{tsp0d3phase}]. One can see that only when $t_s^Y=0$ the magnetic phase diagram from slave-boson mean-field theory agrees well with magnetic orders determined by RKKY interaction via perturbation in weak hybridization regime. In the strong hybridization regime, the effects of small $t_s^Y$ on magnetism can be neglected compared to the large $t_{sp}$. Then magnetic phase diagrams with $t_s^Y=0.1$ and $t_s^Y=0$ are nearly the same.

The transition from PKM or topological PKI phase to trivial PKI phase, characterized by the gapless PKM line in [Fig.~\ref{phase} a)-b)], occurs when the renormalized $s$ orbital on-site energy increases so that $\beta_0-\Delta _s$ satisfies $\beta_0-\Delta _s>|z_+^2t_s^Y+t_p^X|$. We note that the PKM has a vanishing indirect gap, but still has direct gap which is defined as the minimal energy difference of the upper and lower band states at fixed momentum (Fig.~\ref{band}). However, as the energy minimum of the upper subbands equals to the energy maximum of the lower subbands, the Chern number of the PKM phase still denotes the topological invariant of the entire lower subbands.

For the paramagnetic Kondo phases obtained from slave-boson mean-field theory, the $s$ orbital on-site energy $-\Delta_s$ is renormalized from far below $0$ to above $0$ (the $p_X$ orbital on-site energy is $0$) and increases with the hybridization $t_{sp}$ (see Fig.~\ref{moment_gap} a)), and similar results was also shown by Ref.~\cite{mott}. In weak bybridization limit, such property can be verified by the analytic solution provided in Refs.~\cite{dmrg, newns}, where $\beta_0-\Delta_s\propto e^2$ is given. In strong $t_{sp}$ regime, such property in our result can be understood in the following way. In the paramagnetic phase, the mean-fields $z_-=0,\beta=0,p=0$ and the number of order parameters can be reduced to two, i.e. $z_+$ and $\beta_0$, because $e$ and $p_0$ can be viewed as functions of $z_+$ from Eq.~\eqref{constraints} and Eq.~\eqref{zplus}:
 \begin{equation}
  \begin{aligned}
&e=\frac{z_+}{\sqrt{2-z_+^2}},\\
&p_0=\sqrt{1-e^2}.
 \end{aligned}
 \label{e_and_z}
 \end{equation}
  The Hamiltonian matrix can also be reduced to
 \begin{equation}
\epsilon_{k}  =
  \left(
  \begin{matrix}
  z_+^2\epsilon_{s\bold k}+\beta_0 & z_+V_{\bold k}\\
  z_+V_{\bold k}&\epsilon_{p,\bold k}
  \end{matrix}
  \right).
  \label{paramatrix}
  \end{equation}

From the formulas in Eq.~\eqref{constraints} one can see that $e_i^2+\sum_{\mu}\hat{p}_{i\mu}^{\dagger}\hat{p}_{i\mu}=e_i^2+\sum_{\sigma}c_{i\sigma}^{\dagger}c_{i\sigma}=1$ and the $s$ orbital occupation number equals to the pseudo-fermion occupation number since $d_i=0$.  As $t_{sp}$ increases, the renormalized hybridization $z_+ t_{sp}$ and the holon number $e^2$ increases (see Fig.~\ref{moment_gap} a)), so the pseudo-fermion occupation number $\sum_{\sigma}c_{i\sigma}^{\dagger}c_{i\sigma}$ decreases. As a result, one can deduce $\beta_0$ should increase. Otherwise, if the renormalized on-site energy $\beta_0-\Delta_s$ decreases or keeps unchanged, with the increase of $z_+ t_{sp}$, the pseudo-fermion occupation number $\sum_{\sigma}c_{i\sigma}^{\dagger}c_{i\sigma}$ in the fully filled lower band will increase towards $1/2$ for each spin component, since the hybridization in the Hamiltonian matrix Eq.~\eqref{paramatrix} is off-diagonal term and two eigenvectors of $\epsilon_k$ will approach $[1,1]^T/\sqrt{2}$ and $[1,-1]^T/\sqrt{2}$ in large $t_{sp}$ limit.

The PKM to PKI transition that occurs at $\beta_0-\Delta _s=2(z_+^2t_s^Y+t_p^X)$ can be identified by looking at special $k$ points. Near the transition point, the minimum energy of the upper paramagnetic band $E^+$ and maximum energy of the lower paramagnetic band $E^-$ lie within high symmetry $k$ points $(\pi,0)$ and $(0,\pi)$. For the $k$ point $(\pi,0)$, energies of upper and lower bands are
\begin{equation}
 \begin{aligned}
&E^+(\pi,0)=2t_s^Y+\beta_0-\Delta_s,\\
&E^-(\pi,0)=-2t_p^X.
\end{aligned}
\end{equation}
For the $k$ point $(\pi,0)$, however, when $2(z_+^2t_s^Y+t_p^X)< \beta_0-\Delta _s$, the energies of upper and lower  bands are
\begin{equation}
  \begin{aligned}
&E^+(0,\pi)=-2t_s^Y+\beta_0-\Delta_s,\\
&E^-(0,\pi)=2t_p^X,
 \end{aligned}
 \end{equation}
 while when $ \beta_0-\Delta _s< 2(z_+^2t_s^Y+t_p^X)$, the energies of upper and lower  bands are
 \begin{equation}
  \begin{aligned}
&E^+(0,\pi)=2t_p^X,\\
&E^-(0,\pi)=-2t_s^Y+\beta_0-\Delta_s.
 \end{aligned}
 \end{equation}
One can see from the above energies that when $t_s^Y=0$, the paramagnetic phase will evolve from metal to insulator at the transition point, while when $t_s^Y>0$, the paramagnetic phase is always insulator before and after the transition.

In the AFKI phase, the staggered magnetic moment $m_s$ on $s$ orbitals is plotted in Fig.~\ref{moment_gap} b). The $m_s$ approaches local moment limit $m_s\rightarrow 1$ as $t_{sp}\rightarrow 0$, and decreases with the $t_{sp}$ increasing until a continuous transition to paramagnetic phase occurs. The magnetic moment $m_p$ on $p_X$ orbitals is always zero in AFKI or CAFKI phase, for in our checkerboard superlattice each $p_X$ orbital's four nearest $s$ orbitals are in AF or CAF order. The effective Kondo interaction between $s$ and $p_X$ orbitals is spin-spin interaction, so particles on $p_X$ orbitals experience frustrated effective magnetic field from $s$ orbitals and thus have no magnetic moment.

\subsection{QAH phase diagram}
The QAH phase diagrams with $(t_p^X,\Delta _s)=(1,3)$, $t_s^Y=0$ and $0.1$ are plotted in Fig.~\ref{phase} a)-b) respectively. With such parameters, in the single particle regime without interaction, the QAH effect is trivial with $\text{Ch}_1=0$ since $|\Delta_s|>2|t_s^Y+t_p^X|$~\cite{coldatom3}. However, in our CK model $U_s\rightarrow + \infty$ and with the slave-boson mean-field theory, the renormalization for $s$ orbital on-site energy which raises $-\Delta_s$ to $\beta_0-\Delta_s$ may cause band crossing between $s$ and $p_X$ orbitals and thus possibly make the QAH phase non-trivial. In weak hybridization regime, as shown in Fig.~\ref{phase}, both AFKI and CAFKI phases are fully gapped and have quantized Chern number $ \text{Ch}_1=2$. When $t_{sp}$ gradually increases, the CAFKI phase near $\phi=\pi$ evolves to gapless CAFKM phase and FKM phase with unquantized Chern number, while the CAFKI phase near $\phi=\pi$ and AFKI phase around $\phi=\pi/2$ are always gapped with $ \text{Ch}_1=2$. For the PKM or PKI phases in the large $t_{sp}$ regime and on the left side of the PKM line, $|\beta_0-\Delta_s|<2|z_+^2t_s^Y+t_p^X|$ is satisfied and total Chern number $\text{Ch}_1=2$ with each spin component having Chern number $1$. The PKM phase with $t_s^Y=0$ is at the edge between gapless phase and gapped phase as discussed in last subsection, for the maximum of the lower band equals to the minimum of the higher band, so the PKM phase has quantized $\text{Ch}_1$ although it's a metallic phase. On the right side of PKM line, the gap opens again with $|\beta_0-\Delta_s|>2|t_s^Y+t_p^X|$ and QAH effect is trivial. The phases on the special line where $\phi=0$ or $\pi$ also have trivial band topology, because the hybridization function $V_{\bold k}$ is real and the berry curvature vanishes.

\begin{figure}[h]
\centering
\includegraphics[width=3.3in]{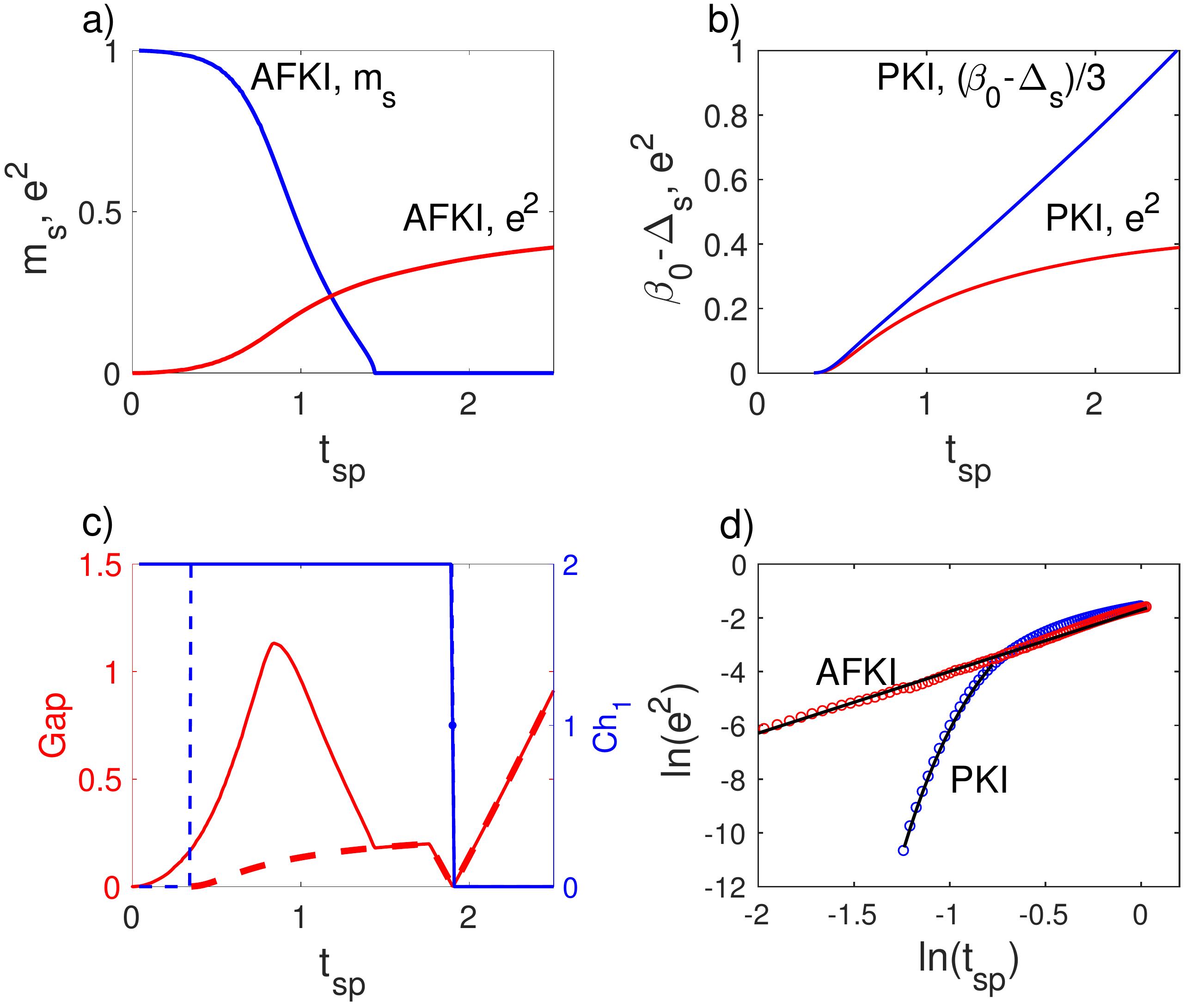}
\caption{Numerical results for the parameter condition $(t_s^Y,t_p^X,\Delta_s,\phi)=(0.1,1,3,0.6\pi)$, computed with slave-boson theory. a) The renormalized $s$-orbital on-site energy $\beta_0-\Delta_s$ and holon number $e^2$ for PKI phase. b) The staggered $s$-orbital magnetic moment $m_s$ and holon number $e^2$ for the AFKI phase. c) Quasiparticle gap and Chern number $\text{Ch}_1$ for the AFKI phase (solid lines) and PKI phase (dashed lines). The Chern number for both phases are different only in weak $t_{sp}$ regime. In the PKI phase, it shows $e^2=0$ when $\ln{(t_{sp})}<-1.23$, and then both the quasiparticle gap and $\text{Ch}_1$ vanishes. d) The scaling $\ln{(e^2)}$ versus $\ln{(t_{sp})}$ for PKI/AFKI phase (circles) and their fitted curves (solid lines) in weak $t_{sp}$ regime. For PKI phase, the fitted curve reads $e^2\propto t_{sp}^{-2}\exp(-1.078/t_{sp}^2)$ which ends for $\ln{(t_{sp})}<-1.23$, while for AFKI phase, the fitted line reads $e^2\propto t_{sp}^{2.29}$ which is nonzero for finite $t_{sp}$. }
\label{moment_gap}
 \end{figure}

We now investigate the effects of magnetism on QAH effect. The quasiparticle gap of the QAH phase opens due to the strongly renormalized $s$-$p_X$ hybridization. In the Coleman slave-boson representation, the hybridization $t_{sp}$ is renormalized in the form of $et_{sp}$, indicating that the non-zero effective hybridization is achieved with $e^2>0$ local moment for each $s$ orbital. In the paramagnetic phase in KR slave-boson representation, $e$ is just a monotonically increasing function of $t_{sp}$'s renormalization factor $z_+$ as shown by Eq.~\eqref{zplus}, so the above physical pictures also applies to the KR slave-boson representation. Fig.~\ref{moment_gap} a) shows numerical results of the $s$ orbital holon number $e^2$ and renormalized on-site energy of the PKI phase with $(t_s^Y,t_p^X,\Delta_s,\phi)=(0.1,1,3,0.6\pi)$. Below the critical hybridization $\ln{(t_{sp}^c)}\approx-1.23$, the holon number has a transition to $e^2=0$, representing the $s$ and $p_X$ orbitals are decoupled and gap closes (Fig.~\ref{moment_gap} c)), i.e. the CK transition described in the previous work~\cite{chernkondo}. The analytic result for paramagnetic slave-boson solutions for standard periodic Anderson model~\cite{dmrg,green} shows exponential relation $e^2\propto J_K^{-1}\exp{(1/J_K\rho)}$ in the $t_{sp}\rightarrow0$ limit, where the Kondo coupling satisfies $J_K\propto t_{sp}^2$ and constant density of state for itinerant band has been assumed in their derivation. Although our CK model is anisotropic, the exponential relation $e^2\propto J_K^{-1}\exp{(1/J_K\rho)}$ in the  PKI phase can be fitted well when $t_{sp}>t_{sp}^c$ as shown in Fig.~\ref{moment_gap} d). The sudden decrease to zero of $e$ in our numeric solution of the mean-field equations when $t_{sp}<t_{sp}^c$ may come from the inaccuracy of the numerical calculation, for $e^2$ decreases towards $0$ so fast in weak $t_{sp}$ limit. However, since $e^2$ decreases so fast we can still view the $t_{sp}^c$ as a quasi-critical point, and below this $t_{sp}^c$ the renormalized hybridization can be regarded as $0$ and the $\text{Ch}_1$ vanishes.

Compared to the PKI phase, in the magnetic Kondo phase the holon number $e^2$ does not decrease so fast and effective hybridization is enhanced in weak hybridization limit. Among the magnetic Kondo phases, we found in our numerical results there is no much difference in the holon number $e^2$ between different types of magnetic orders. We plot holon nummber $e^2$ of the AFKI phase in Fig.~\ref{moment_gap} b) as an example, where $e^2$ is far more larger than that of the PKI phase in weak hybridization limit, implying the effective hybridization is enhanced compared to paramagnetic phase in such limit. Similar results was also shown by Refs.~\cite{sbmf,anderson} for standard Anderson model. Another numerical study of standard one-dimensional periodic Anderson model with density-matrix renormalization group (DMRG)~\cite{dmrg} found that the relation between $e^2$ and $t_{sp}$ is power law $ \ln{e}\propto\ln{ t_{sp}}$ at weak $t_{sp}$ with infinite large $U_s$ in Kondo regime, in contrast to the PKI phase obtained with slave-boson mean-field theory, where $e^2\propto J_K^{-1}\exp{(1/J_K\rho)}$. It was also found that although the ground state has zero total spin, the antiferromagnetic correlation is strong. For the present CK model, we also plotted the fitted power-law relation $ \ln{e} \propto \ln{t_{sp}}$ for the AFKI phase in Fig.~\ref{moment_gap} d), which qualitatively agrees with the DMRG result in Ref.~\cite{dmrg}. As the magnetic Kondo phases is energetically more stable than the PKI phase in weak hybridization regime, the CK transition in the PKI phase described in the previous work~\cite{chernkondo} will not occur in our CK model.

The present mean-field calculation also shows that different magnetic orders have different influences on QAH effect. The AF order always enhance the quasiparticle gap of the QAH phase compared to the paramagnetic phase, and we plot the comparison of gap between AFKI and PKI phase with $\phi=0.6\pi$ in Fig.~\ref{moment_gap} c) as an example. In Fig.~\ref{moment_gap} c), the gap of the AFKI phase has a peak for $t_{sp}\approx 0.8$ that results from the increase of the self-consistent magnetic field $\beta$ on $s$ orbital~\cite{sbmf, dorin}. Note that other magnetic orders do not always enhance the gap. For CAF order, we obtain a CAFKI phase in weak hybridization regime where the quasiparticle gap is enhanced, but as the hybridization increases, we obtain a CAFKI phase near $\phi\gtrsim 0$ and obtain a CAFKM phase near $\phi\approx \pi$ as shown from Fig.~\ref{phase} with moderate $t_{sp}$. For the CAFKM phase, the quasiparticle is gapless and the Hall effect is not quantized any more. For ferromagnetic order, we always obtain a gapless FKM phase since the spin splitting is uniform in real space and thus the Hall effect is always not quantized.

 \begin{figure}[h]
\centering
\includegraphics[width=3in]{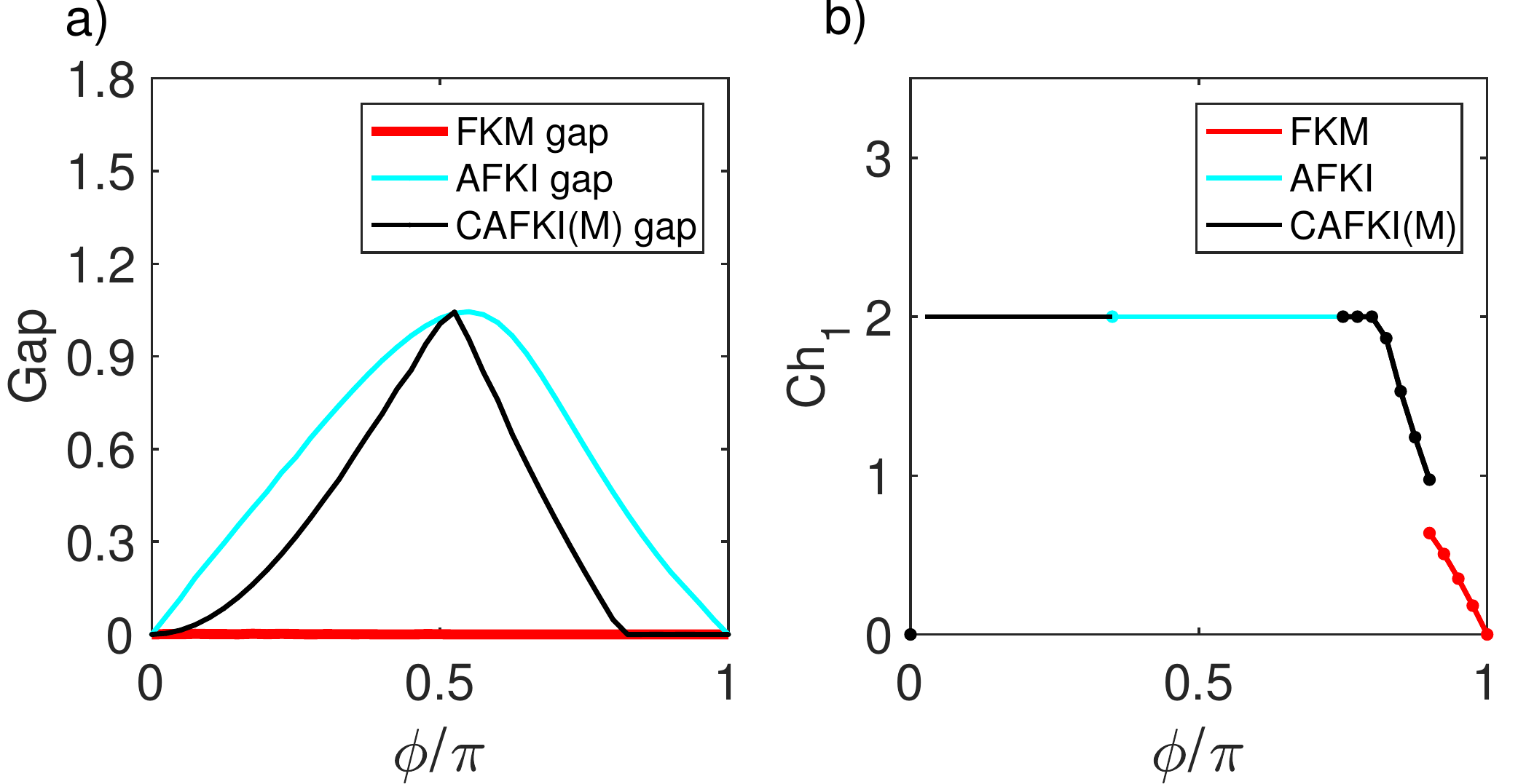}
\caption{Quasiparticle gaps, magnetic orders and Chern number $\text{Ch}_1$ with $(t_s^Y,t_p^X,t_{sp},\Delta_s)=(0.1,1,0.77,3)$ obtained with slave-boson theory. a) Quasiparticle gaps for three types of magnetic orders versus $\phi$. b) Ground state magnetic order and first Chern number $\text{Ch}_1$ as a function of $\phi$.  }
\label{tsp0d77}
 \end{figure}

\subsection{Experimental measurement of the topology and strong correlation effects}

The fully controllable cold atom experimental technologies including the Hall effect measurement~\cite{bandtopology} and double occupancy measurement~\cite{measuredouble1, measuredouble2} can enable us to identify the topology and influences of strong correlation on the CK phase. We now predict and discuss the observables including Hall conductance and double occupancy that can be affected by the topology, correlation effect respectively. With the coexistence of magnetic order, these observables in magnetic CK phase will be qualitatively different from that in PKI phase.

The non-trivial band topology is determined by the existence of quasiparticle gap and band inversion. As having been discussed in the last subsection, the quasiparticle gap is affected by the existence of effective Kondo hybridization and the magnetic orders, while the band inversion is affected by the magnitude of the renormalized $s$-$p_X$ on-site energy difference, which is controlled by the strength of $t_{sp}$. This leads to a rich QAH phase diagram. To identify the topological physics, one can either tune the magnitude of $t_{sp}$ with phase $\phi$ fixed, or tune phase $\phi$ with the magnitude of $t_{sp}$ fixed. In particular, one can tune the magnitude of $t_{sp}$ with parameters $(t_s^Y,t_p^X,\Delta_s,\phi)=(0.1,1,3,0.6\pi)$ being fixed. In the noninteracting regime the $s$ orbital on-site energy lies far below $p_X$ orbital, and the phase is trivial irrespective of magnitude of $t_{sp}$. However, in the presence of strong repulsive interaction, the $s$-$p_X$ on-site energy difference is strongly renormalized to a small quantity, with the effective hybridization being enhanced by the AF magnetic order in contrast to the paramagnetic phase, leading to the non-trivial QAH effect with $\text{Ch}_1=2$ when $t_{sp}$ is not too strong, as shown in Fig.~\ref{moment_gap} c). We note that for large enough $t_{sp}$, the on-site energy difference can be further renormalized, finally yielding a large magnitude again, and the phase can reenter the trivial regime, as shown in Fig.~\ref{moment_gap} b) and c). Further, one can also tune $\phi$ from $0$ to $\pi$ and keep $(t_s^Y,t_p^X,t_{sp},\Delta_s)=(0.1,1,0.77,3)$ being fixed. In this case, the ground state phases can be CAFKI, AFKI, CAFKI(M), or FKM phases with different $\phi$ [Fig.~\ref{phase} and Fig.~\ref{tsp0d77}]. Among these magnetic orders, the AF order always lead to an insulating phase with enhanced gap and QAH effect, the CAF order may lead to insulating or metallic phase determined by $t_{sp}$ and $\phi$, while the ferromagnetic order always result in metallic phase without quantized Hall effect.
\begin{figure}[t]
\centering
\includegraphics[width=3in]{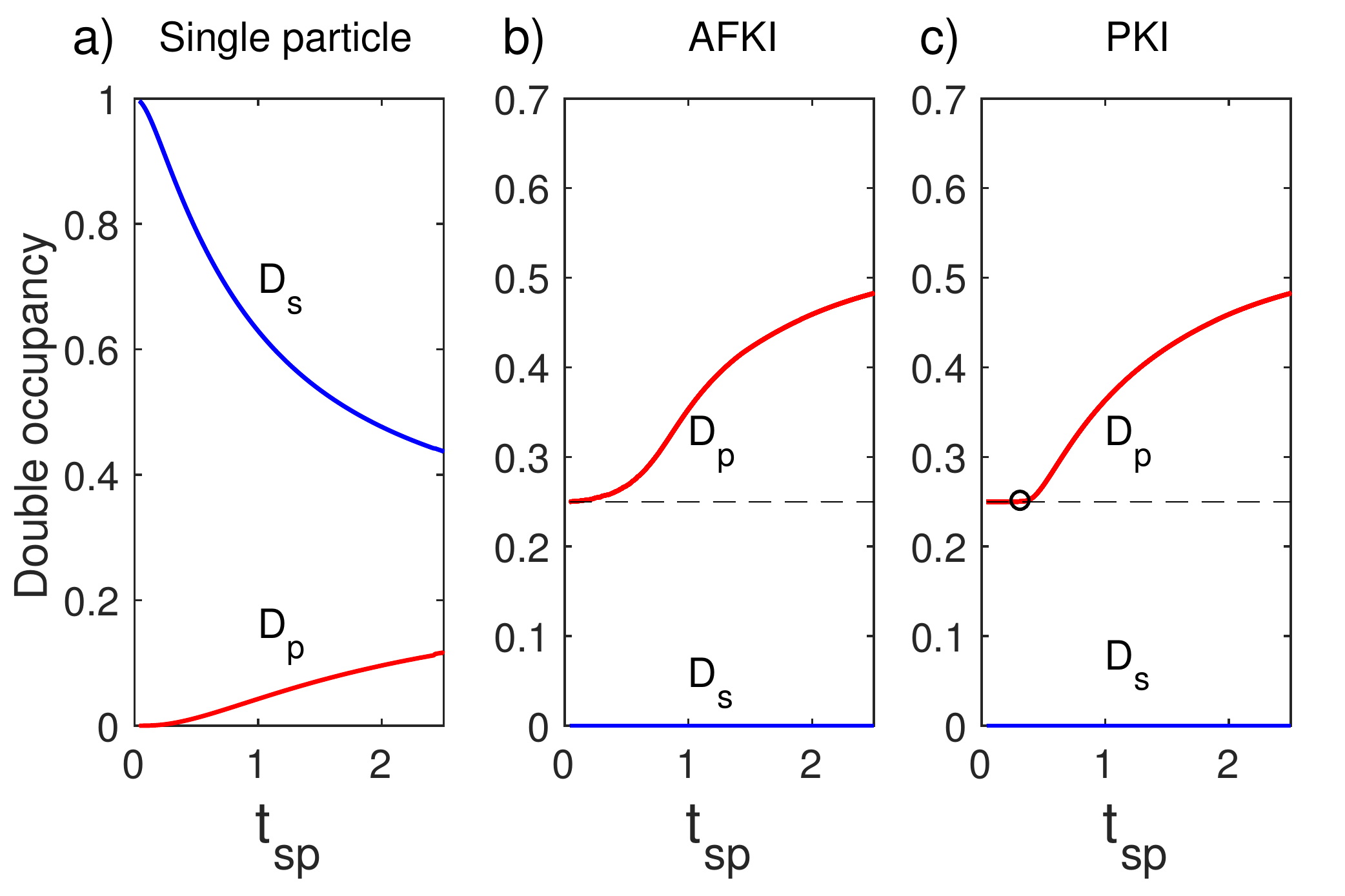}
\caption{Double occupancy probability for $s$ orbitals ($D_s$) and for $p_X$ orbitals ($D_p$) with $(t_s^Y,t_p^X,\Delta_s,\phi)=(0.1,1,3,0.6\pi)$. a) Single particle regime with $U_s=0$. b) AFKI phase in the regime with $U_s\rightarrow + \infty$. c) PKI in the regime with $U_s\rightarrow +\infty$. The results in b)-c) are obtained with slave-boson theory. Note that $D_p$ in c) has a transition point labeled by a black circle. As seen in Fig.~\ref{moment_gap} a), when $t_{sp}\gtrapprox1.45$, the AFKI magnetic moment vanishes and the system enters the PKI phase.}
\label{occupy}
 \end{figure}

Concerning the strong correlation effects, we show the double occupancy $D_p$ for $p_X$ orbital and $D_s$ for $s$ orbital with $(t_s^Y,t_p^X,\Delta_s,\phi)=(0.1,1,3,0.6\pi)$ in Fig.~\ref{occupy} b). We also calculated the double occupancy for single particle regime ($U_s=0$) and PKI phase as a comparison in Fig.~\ref{occupy} a) and c). The subfigure a) differs from b) and c) for lack of correlation effect, while b) differs from c) without consideration of the magnetism. The $D_p$ for non-interaction $p_X$ orbital is calculated by Wick's theorem $D_p=n_{p\uparrow}n_{p\downarrow}$ for both a), b), and c). In the single particle regime with $U_s=0$ corresponding to subfigure a), the on-site energy for $s$ orbital lies below that of $p_X$ orbital and thus the particle number and double occupancy on $s$ orbitals is larger than that on $p_X$ orbitals. Fig.~\ref{occupy} b)-c) correspond to AFKI phase and PKI phase respectively in strong correlated regime with repulsive $U_s\rightarrow\infty$. Before the laser assisted hybridization $t_{sp}$ is induced, the $s$ orbitals form a half filled Mott insulator and the double occupancy $D_s$ on $s$ orbitals is suppressed to zero with the $s$ orbital on-site energy being renormalized to above $p_X$ orbital due to strong $U_s$. Thus in b) and c) $D_p$ is greatly enhanced compared to that in single particle regime in Fig.~\ref{occupy} a). When hybridization $t_{sp}$ is induced and effective hybridization exists, $D_s$ keeps to be zero, while particles on $s$ orbitals begin to pump into $p_X$ orbitals, with the $s$ orbital on-site energy being further renormalized. The difference between Fig.~\ref{occupy} b) and Fig.~\ref{occupy} c) occurs in weak hybridization $t_{sp}$ limit. For the AFKI case with magnetic order, effective hybridization is enhanced and exists as long as $t_{sp}\neq 0$, while for PKI phase the existence of effective hybridization needs the hybridization exceed its quasi-critical value $t_{sp}^c$, i.e., after the CK transition. Around the quasi-critical $t_{sp}^c$, for AFKI phase the $D_p$ increases smoothly while for the PKI phase the $D_p$ starts to increase abruptly at $t_{sp}\gtrsim t_{sp}^c$, showing the difference between emergence of the magnetic and paramagnetic CK phases.

\section{Conclusions}

In this work, we have examined the Chern Kondo insulator by revisiting its realization and studied the magnetic effects on the Chern Kondo phases. An improved scheme for the realization of Chen Kondo insulator is proposed, solving the challenges in the previous realization. The Ruderman-Kittel-Kasuya-Yoshida magnetic interaction is analyzed at weak hybridization limit, with the anisotropic magnetic effects being discussed. We further systematically studied the paramagnetic and magnetic phases coexisting with Kondo hybridization based on slave-boson theory and mapped out the full magnetic and correlated QAH phase diagrams. The rich phases, including the paramagnetic/magnetic Kondo insulating phases and magnetic Kondo metallic phases, have been obtained and investigated in detail. Interestingly, the effective Kondo hybridization can be typically strengthened by taking into account magnetic effects. In particular, we showed that the existence of antiferromagnetic order enhances the Kondo phase, with the topological bulk gap being increased compared with that in the paramagnetic regime. On the other hand, the kondo phases coexisting with collinear antiferromagnetic order have metal-insulator transition determined by the strength and phase of hybridization, which is absent in the paramagnetic Kondo phase. Moreover, in the large hybridization regime, the bulk phase may eventually enter the paramagnetic Kondo insulating states, which manifests that the magnetic orders are fully suppressed in the strongly Kondo regime. The Chern Kondo phases can be detected by measuring the Chern number of bulk topology and the double occupancy, which are achievable in cold atom experiments. The rich strongly correlated and topological physics may motivate further studies of the Chern Kondo phases in theory and experiment.

\section{acknowledgement}

We appreciate the discussion with Hua Chen. This work was supported by the National Key R\&D Program of China (2016YFA0301604), National Nature Science Foundation of China (under grants N0. 11574008 and No. 11761161003), and the Thousand-Young-Talent Program of China.

   \noindent

\onecolumngrid

\renewcommand{\thesection}{S-\arabic{section}}
\setcounter{section}{0}  
\renewcommand{\theequation}{S\arabic{equation}}
\setcounter{equation}{0}  
\renewcommand{\thefigure}{S\arabic{figure}}
\setcounter{figure}{0}  

\indent

\section*{\Large\bf APPENDIX}

\subsection{The $s$-$p_X$ hopping integrals from laser assisted tunneling in the previous realization}

In this subsection we will evaluate the hopping integrals for original laser assisted tunneling in Fig.~\ref{tunnel} a):
   \begin{equation}
  \begin{aligned}
&J_1=\int d^2r\psi_{n,m}^p(x,y)\psi_{n+1,m}^s(x,y)e^{ik_R(y-m)}e^{ik_Rm},\\
&J_2=\int d^2r\psi_{n+1,m+1}^p(x,y)\psi_{n+1,m}^s(x,y)e^{-ik_R(y-(m+\frac{1}{2}))}e^{-ik_R(m+\frac{1}{2})},\\
&J_3=\int d^2r\psi_{n+1,m+1}^p(x,y)\psi_{n,m+1}^s(x,y)e^{ik_R(y-(m+1))}e^{ik_R(m+1)},\\
&J_4=\int d^2r\psi_{n,m}^p(x,y)\psi_{n,m+1}^s(x,y)e^{-ik_R(y-(m+\frac{1}{2}))}e^{-ik_R(m+\frac{1}{2})}.
  \end{aligned}
  \end{equation}
Here $\psi_{n,m}^p(x,y)$ and $\psi_{n,m}^s(x,y)$ are real maximally localized Wannier functions for $p_X$ ($s$) orbitals, $(n,m)$ is the coordinate of lattice site and $\psi_{0,0}^p(x,y)$/$\psi_{0,0}^s(x,y)$ are odd/even functions respectively. We also have $\psi_{n,m}(x,y)=\psi_{0,0}(x-n,y-m)$, then the integrals can be simplified:
   \begin{equation}
  \begin{aligned}
I_1&=\int d^2r\psi_{n,m}^p(x,y)\psi_{n+1,m}^s(x,y)e^{ik_R(y-m)}\\&=\int d^2r\psi_{0,0}^p(x,y)\psi_{1,0}^s(x,y)e^{ik_Ry},\\
I_3&= \int d^2r\psi_{n+1,m+1}^p(x,y)\psi_{n,m+1}^s(x,y)e^{ik_R(y-(m+1))}\\&= \int d^2r\psi_{1,0}^p(x,y)\psi_{0,0}^s(x,y)e^{ik_Ry}\\
&=\int d^2r\psi_{0,0}^p(x-1,y)\psi_{0,0}^s(x,y)e^{ik_Ry}\\&=-\int d^2r\psi_{0,0}^p(-x+1,-y)\psi_{0,0}^s(-x,-y)e^{ik_Ry}\\
&=-\int d^2r\psi_{0,0}^p(x+1,y)\psi_{0,0}^s(x,y)e^{-ik_Ry}\\&=-\int d^2r\psi_{0,0}^p(x,y)\psi_{1,0}^s(x,y)e^{-ik_Ry}.
  \end{aligned}
  \end{equation}
  In the above formulas we have used the property of integral:
   \begin{equation}
  \begin{aligned}
  \int dr^2f(-x,-y)=\int dr^2rf(x,y).
    \end{aligned}
  \end{equation}
  Decomposing the complex integral into real and imaginary parts, we can obtain the relation between $I_1$ and $I_3$:
  \begin{equation}
  \begin{aligned}
I_1=&+\int d^2r\psi_{0,0}^p(x,y)\psi_{1,0}^s(x,y)\cos(k_Ry)+i\int d^2r\psi_{0,0}^p(x,y)\psi_{1,0}^s(x,y)\sin(k_Ry)=I_a,\\
I_3=&-\int d^2r\psi_{0,0}^p(x,y)\psi_{1,0}^s(x,y)\cos(k_Ry)+i\int d^2r\psi_{0,0}^p(x,y)\psi_{1,0}^s(x,y)\sin(k_Ry)=-I_a^*.
  \end{aligned}
  \end{equation}
Similarly,
 \begin{equation}
   \begin{aligned}
I_2&=\int d^2r\psi_{n+1,m+1}^p(x,y)\psi_{n+1,m}^s(x,y)e^{-ik_R(y-(m+\frac{1}{2}))}\\&= \int d^2r\psi_{0,0}^p(x,y-\frac{1}{2})\psi_{0,0}^s(x,y+\frac{1}{2})e^{-ik_Ry}\\
&=-\int d^2r\psi_{0,0}^p(-x,-y+\frac{1}{2})\psi_{0,0}^s(-x,-y-\frac{1}{2})e^{-ik_Ry}\\&=-\int d^2r\psi_{0,0}^p(x,y+\frac{1}{2})\psi_{0,0}^s(x,y-\frac{1}{2})e^{ik_Ry},\\
I_4&=\int d^2r\psi_{n,m}^p(x,y)\psi_{n,m+1}^s(x,y)e^{-ik_R(y-(m+\frac{1}{2}))}\\&=\int d^2r\psi_{0,0}^p(x-n,y-m)\psi_{0,0}^s(x-n,y-m-1)e^{-ik_R(y-(m+\frac{1}{2}))}\\
&=\int d^2r\psi_{0,0}^p(x,y+\frac{1}{2})\psi_{0,0}^s(x,y-\frac{1}{2})e^{-ik_Ry}.
  \end{aligned}
  \end{equation}

 Decomposing these two number into real and imaginary part, we can obtain the relation between $I_2$ and $I_4$:
  \begin{equation}
  \begin{aligned}
I_2=&-\int d^2r\psi_{0,0}^p(x,y+\frac{1}{2})\psi_{0,0}^s(x,y-\frac{1}{2})\cos(k_Ry)-i\int d^2r\psi_{0,0}^p(x,y+\frac{1}{2})\psi_{0,0}^s(x,y-\frac{1}{2})\sin(k_Ry)=-I_b^*,\\
I_4=&+\int d^2r\psi_{0,0}^p(x,y+\frac{1}{2})\psi_{0,0}^s(x,y-\frac{1}{2})\cos(k_Ry)-i\int d^2r\psi_{0,0}^p(x,y+\frac{1}{2})\psi_{0,0}^s(x,y-\frac{1}{2})\sin(k_Ry)=I_b.
  \end{aligned}
  \end{equation}

\subsection{Effective Kondo lattice Hamiltonian}
The effective Kondo lattice Hamiltonian is derived through perturbation theory when $s$-$p_X$ hybridization is weak and the $s$ orbital on-site energy lies far below the $p_X$ orbital. Here we provide detailed derivation of the Kondo lattice Hamiltonian
   \begin{equation}
    \begin{aligned}
H_{KL}= \sum\limits_{\bold k\sigma}\epsilon_{p\bold k} p_{X\bold k\sigma}^{\dagger}p_{X\bold k\sigma}+\sum\limits_{i,\bold k,\bold k'}J_{\bold k,\bold k',i}\bold{S}_i\cdot \bold{s}_{\bold k \bold k'},
   \end{aligned}
  \end{equation}
  from the original Hamiltonian
  \begin{equation}
    \begin{aligned}
    H=&H_1+H',\\
H_1=&\sum\limits_{i\sigma}\left[ -\Delta_ss_{i\sigma}^{\dagger}s_{i\sigma}+t_{p}^Xp_{Xi\sigma}^{\dagger} p_{Xi\pm \hat{X}\sigma} \right]\\
&+\sum\limits_{i}U_s\hat{n}_{si\uparrow}\hat{n}_{si\downarrow},\\
H'=&\sum\limits_{\bold k,i}\frac{V_{\bold k}e^{-i\bold k\cdot \bold R_i}}{\sqrt{N}}s_{i\sigma}^{\dagger}p_{Xk\sigma}+\mathrm{H.c.},
  \end{aligned}
  \end{equation}
  and the definition of the effective Hamiltonian
   \begin{equation}
    \begin{aligned}
&H_p(E)=PHP-PHQ\frac{1}{QHQ-E} QHP.
   \end{aligned}
  \end{equation}
  where the projection operator $P$ project states onto subspace with each $s$ orbital singly occupied and $Q=1-P$.

We separate $H'$ into $H'=H_++H_-$, where $H_+=\sum_{\bold k,i}N^{-1/2}V_{\bold k}e^{-i\bold k\cdot \bold R_i}s_{i\sigma}^{\dagger}p_{Xk\sigma}$ and $H_-=\sum_{\bold k,i}N^{-1/2}V_{\bold k}e^{i\bold k\cdot \bold R_i}p_{Xk\sigma}^{\dagger}s_{i\sigma}$. The operator $H_+$ increases one particle on $s$ orbitals and $H_-$ decreases one particle on $s$ orbitals.

Making the approximation that replacing the unknown $E$ by the unperturbed energy $E_0$, and with the formulas
   \begin{equation}
    \begin{aligned}
&PH_1Q=0,\\
&PH'P=0,\\
&QH'Q=0,
   \end{aligned}
  \end{equation}
the effective Hamiltonian can be simplified to the form
     \begin{equation}
    \begin{aligned}
&H_p(E)\approx P(H_1+H_++H_-)P-P(H_1+H_++H_-)Q\frac{1}{QHQ-E_0} Q(H_1+H_++H_-)P\\
=&PH_1P-PH_+Q\frac{1}{QH_1Q-E_0} QH_-P-PH_-Q\frac{1}{QH_1Q-E_0} QH_+P\\
=&PH_1P-PH_+Q\frac{1}{QH_1Q-E_0} QH_-P-PH_-Q\frac{1}{QH_1Q-E_0} QH_+P.\\
   \end{aligned}
  \end{equation}

Substituting $H_1$, $H_+$ and $H_-$ with their definitions, $H_p$ takes the form
     \begin{equation}
    \begin{aligned}
&H_p(E)=\sum\limits_{\bold k\sigma}\epsilon_{p\bold k} p_{X\bold k\sigma}^{\dagger}p_{X\bold k\sigma}-\sum_{\substack{\bold k,\bold k'\\i,\sigma,\sigma'}} \frac{V_{\bold k}^{*} V_{\bold k^{\prime}}e^{i(\bold k -\bold k')\cdot \bold R_i}}{N}\frac{s_{i\sigma^{\prime}}^{\dagger}p_{X\bold k^{\prime}\sigma^{\prime}}p_{X\bold k\sigma}^{\dagger}s_{i\sigma}}{\epsilon_{p\bold k}+\Delta_s}-\sum_{\substack{\bold k,\bold k'\\i,\sigma,\sigma'}} \frac{V_{\bold k}^{*} V_{\bold k^{\prime}}e^{i(\bold k' -\bold k)\cdot \bold R_i}}{N}\frac{p_{X\bold k^{\prime}\sigma^{\prime}}^{\dagger}s_{i\sigma^{\prime}}s_{i\sigma}^{\dagger}p_{X\bold k\sigma}}{U_s-\epsilon_{p\bold k}-\Delta_s}.\\
   \end{aligned}
  \end{equation}

Here we have discarded the on-site energy $-N\Delta_s$ for $s$ orbitals, for it's a constant in the subspace P. The third term in the above formula can be omitted since we only consider the infinitely large $U_s$ limit. Using the identity $s_{i\sigma'}^{\dagger}p_{Xk'\sigma'}p_{Xk\sigma}^{\dagger}s_{i\sigma}=
s_{i\sigma'}^{\dagger}s_{i\sigma}(\delta_{\sigma\sigma'}\delta_{kk'}-p_{Xk\sigma}^{\dagger}p_{Xk'\sigma'})$, the effective Hamiltonian reads

     \begin{equation}
    \begin{aligned}
&H_p(E)=\sum\limits_{\bold k\sigma}\epsilon_{p\bold k} p_{X\bold k\sigma}^{\dagger}p_{X\bold k\sigma}+\sum_{\substack{\bold k,\bold k'\\i,\sigma,\sigma'}} \frac{V_{\bold k}^{*} V_{\bold k^{\prime}}e^{i(\bold k -\bold k')\cdot \bold R_i}}{N}\frac{s_{i\sigma^{\prime}}^{\dagger}s_{i\sigma}p_{X\bold k\sigma}^{\dagger}p_{X\bold k^{\prime}\sigma^{\prime}}}{\epsilon_{p\bold k}+\Delta_s}-\sum_{\substack{\bold k,i,\sigma}} V_{\bold k}^{*} V_{\bold k}\frac{s_{i\sigma}^{\dagger}s_{i\sigma}}{\epsilon_{p\bold k}+\Delta_s}.\\
   \end{aligned}
  \end{equation}

The third term in the above formula is also a constant in the subspace P. Defining the spin density operators $\bold{S}_i=s_{i\sigma^{\prime}}^{\dagger}\boldsymbol \tau_{\sigma ' \sigma}s_{i\sigma}/2$, $\bold{s}_{\bold k,\bold k'}=p_{X\bold k\sigma^{\prime}}^{\dagger}\boldsymbol \tau_{\sigma ' \sigma}p_{X\bold k'\sigma}/2$ where $\boldsymbol \tau$ is the vector formed by three Pauli matrices, we obtain the identities
 \begin{equation*}
    \begin{aligned}
&s_{i\uparrow}^{\dagger}s_{i\downarrow}p_{Xk\downarrow}^{\dagger}p_{Xk'\uparrow}=S_i^{+}s_{kk'}^{-},\\
&s_{i\downarrow}^{\dagger}s_{i\uparrow}p_{Xk\uparrow}^{\dagger}p_{Xk'\downarrow}=S_i^{-}s_{kk'}^{+},\\
&s_{i\uparrow}^{\dagger}s_{i\uparrow}p_{Xk\uparrow}^{\dagger}p_{Xk'\uparrow}+s_{i\downarrow}^{\dagger}s_{i\downarrow}p_{Xk\downarrow}^{\dagger}p_{k'\downarrow}=\frac{1}{2}(p_{Xk\uparrow}^{\dagger}p_{Xk'\uparrow}+p_{Xk\downarrow}^{\dagger}p_{Xk'\downarrow})+S_i^{z}s_{kk'}^{z}.
 \end{aligned}
 \end{equation*}

The potential scattering term $p_{Xk\uparrow}^{\dagger}p_{Xk'\uparrow}+p_{Xk\downarrow}^{\dagger}p_{Xk'\downarrow}$ in the third line  can be omitted if we only care the phenomena about magnetism. Finally we obtain the final effective Kondo lattice Hamiltonian:
 \begin{equation}
 \begin{aligned}
H_{KL}= \sum\limits_{\bold k\sigma}\epsilon_{p\bold k} p_{X\bold k\sigma}^{\dagger}p_{X\bold k\sigma}+\sum\limits_{i,\bold k,\bold k'}2J_{\bold k,\bold k',i}\bold{S}_i\cdot \bold{s}_{\bold k \bold k'}.
  \end{aligned}
  \end{equation}

Here the anisotropic k dependent Kondo coupling $J_{\bold k,\bold k',i}=\frac{1}{N}\frac{V_{\bold k}^{*} V_{\bold k'}e^{i(\bold k-\bold k')\cdot \bold R_i}}{\epsilon_{p\bold k}+\Delta_s}$ contains the information of the hybridization between $s$ and $p_X$ orbitals.

\subsection{The RKKY interaction}
To derive the RKKY interation, we take the $p_X$ orbital hopping terms as unperturbed Hamiltonian and take the Kondo interaction as the perturbation. We seperate the Kondo interaction into three terms, and the Kondo lattice Hamiltonian reads

   \begin{equation*}
    \begin{aligned}
H_{KL}= \sum\limits_{k} \epsilon_{pk} p_{Xk\sigma}^{\dagger}p_{Xk\sigma}+\sum\limits_{k,k',i} J_{k,k',i}(S_i^{-}p_{Xk\uparrow}^{\dagger}p_{Xk'\downarrow}+S_i^{+}p_{Xk\downarrow}^{\dagger}p_{Xk^{\prime}\uparrow}+S_i^{z}\sum\limits_{\sigma} \sigma p_{Xk\sigma}^{\dagger}p_{Xk^{\prime}\sigma}).
   \end{aligned}
  \end{equation*}

Now we define three components of the Kondo interaction $H_-=\sum_{k,k',i} J_{k,k',i}S_i^{-}p_{Xk\uparrow}^{\dagger}p_{Xk'\downarrow}$, $H_+=\sum_{k,k',i} J_{k,k',i}S_i^{+}p_{Xk\downarrow}^{\dagger}p_{Xk^{\prime}\uparrow}$ and $H_z=\sum_{k,k',i} J_{k,k',i}S_i^{z}\sum_{\sigma} \sigma p_{Xk\sigma}^{\dagger}p_{Xk^{\prime}\sigma}$. We also define the projection operator $P_0$ that projects the original Hilbert space of Kondo lattice onto the subspace with a ground state Fermi sea formed by $p_X$ orbital degree of freedom, i.e. $\hat{n}_{pk\sigma}=1$ when $\epsilon_{p,k}<\epsilon_{p,k_F}$, and $\hat{n}_{pk\sigma}=0$ when $\epsilon_{p,k}>\epsilon_{p,k_F}$. The states in subspace $Q_0=1-P_0$ have particle type or hole type excitations in $p_X$ orbital degree of freedom.

Following the steps in the derivation of Kondo lattice Hamiltonian, the effective Hamiltonian $H_{p_0}$ reads
 \begin{equation*}
  \begin{aligned}
H_{p_0}=&P_0H_{KL}P_0+P_0H_zQ_0\frac{1}{Q_0H_{KL}Q-E_0}Q_0H_zP_0\\
&+P_0H_+Q_0\frac{1}{Q_0H_{KL}Q-E_0}Q_0H_-P_0+P_0H_-Q_0\frac{1}{Q_0H_{KL}Q-E_0}Q_0H_+P_0.
 \end{aligned}
 \end{equation*}

The first term, $P_0H_{KL}P_0=\sum_{\epsilon_{pk}<\epsilon_{pk_F}}2\epsilon_{pk}$ is a constant term and doesn't affect the magnetism.
 The second term takes the form
 \begin{equation*}
 \begin{aligned}
&P_0H_zQ_0\frac{1}{Q_0H_{KL}Q-E_0}Q_0H_zP_0=\sum_{kk'ij}J_{kk'i}J_{k'kj}S_i^{z}S_j^{z}(p_{k\uparrow}^{\dagger}p_{k^{\prime}\uparrow}p_{k^{\prime}\uparrow}^{\dagger}p_{k\uparrow}+p_{k\downarrow}^{\dagger}p_{k^{\prime}\downarrow}p_{k^{\prime}\downarrow}^{\dagger}p_{k\downarrow})/(\epsilon_{p,k'}-\epsilon_{p,k})\\
=&\sum_{kk'ij}2J_{kk'i}J_{k'kj}S_i^{z}S_j^{z}[n_{p,k}(1-n_{p,k'})]/(\epsilon_{p,k'}-\epsilon_{p,k})\\
=&\sum\limits_{ijkk'}2J_{kk'i}J_{k'kj}S_i^{z}S_j^{z}[n_{p,k}-n_{p,k'}]/(\epsilon_{p,k'}-\epsilon_{p,k}).\\
 \end{aligned}
 \end{equation*}
Similarly, the third term reads
 \begin{equation*}
 \begin{aligned}
P_0H_+Q_0\frac{1}{Q_0H_{KL}Q-E_0}Q_0H_-P_0=\sum\limits_{ijkk'}2J_{kk'i}J_{k'kj}S_i^{+}S_j^{-}[n_{p,k}-n_{p,k'}]/(\epsilon_{p,k'}-\epsilon_{p,k}),\\
 \end{aligned}
 \end{equation*}
 and the forth term reads
\begin{equation*}
 \begin{aligned}
P_0H_-Q_0\frac{1}{Q_0H_{KL}Q-E_0}Q_0H_+P_0=\sum\limits_{ijkk'}2J_{kk'i}J_{k'kj}S_i^{-}S_j^{+}[n_{p,k}-n_{p,k'}]/(\epsilon_{p,k'}-\epsilon_{p,k}).\\
\end{aligned}
\end{equation*}
  Finally, we obtain the RKKY interaction
\begin{equation*}
\begin{aligned}
H_{RKKY}= \sum\limits_{i,j}2J(X_i-X_j,Y_i-Y_j)\bold{S}_i\cdot \bold{S}_j,
\end{aligned}
\end{equation*}
  where coupling coefficient takes the form
\begin{equation}
\begin{aligned}
J(X_i-X_j,Y_i-Y_j)=&\sum\limits_{k,k'}J_{k,k',j}J_{k',k,j}\\
=&\sum\limits_{\bold k,\bold k'}\frac{4\cos[(\bold k-\bold k')\cdot(\bold R_i-\bold R_j)]}{N^2}|V_{\bold k}|^2|V_{\bold k'}|^2\frac{1}{\epsilon_{p\bold k}+\Delta_s}\frac{1}{\epsilon_{p\bold k'}+\Delta_s}\frac{n_{p,\bold k}-n_{p,\bold k'}}{\epsilon_{p,\bold k}-\epsilon_{p,\bold k'}}.
\end{aligned}
\end{equation}

\subsection{Slave-boson mean-field Hamiltonian}
In this subsection, we first review the spin-rotation invariant slave-boson formulas introduced in Ref.~\cite{rotation} and then provide the derivation of mean-field Hamiltonian of the CK model following~\cite{hubbard,anderson}. The spin-rotation invariant type slave-boson theory~\cite{rotation} is convenient to describe various magnetic orders. Furthermore, it will give better results when considering fluctuations around mean-field solutions~\cite{rotation}, although our treatment is only restricted to mean-field level.

The purpose of the slave-boson mean-field theory is to construct composite particle states and Hamiltonian operator that equivalent to the original states and Hamiltonian, and then take the boson fields to be mean-fields as an approximation. In the spin-rotation invariant slave-boson theory~\cite{rotation}, auxiliary bosonic and fermionic operators are introduced. On the one hand, the slave-boson operators $\hat{e},\hat{d},\hat{p}_0, \hat{\bold p}=(\hat{p}_1,\hat{p}_2,\hat{p}_3)$ that obey bosonic commutation relation are introduced. Here $\hat{e}, \hat{d}$ correspond to hole and doubly occupied states; scalar ($S=0$) field $\hat{p}_0$ and vector ($S=1$) field $\hat{\bold p}=(\hat{p}_1,\hat{p}_2,\hat{p}_3)$ correspond to the singly occupied state. $\hat{e},\hat{d},\hat{p}_0$ transform as a scalar under spin rotation and $\hat{\bold p}$ transforms as a vector. On the other hand, the $S =1/2$ pseudo-fermion operators $c_{i\sigma }$ obey fermionic commutation relation.

The holon and doublon states can be constructed directly
    \begin{equation*}
  \begin{aligned}
|0\rangle&= \hat{e}^{\dagger}|Vac\rangle,\\
|\uparrow\downarrow\rangle&= \hat{d}^{\dagger}c_{\uparrow}^{\dagger}c_{\downarrow}^{\dagger}|Vac\rangle,
    \end{aligned}
  \end{equation*}
where $|\text{Vac}\rangle$ is the vacuum for both boson and fermion states. Concerning the singly occupied states, there are two ways to construct a composite $S=1/2$ states via combining the slave-boson operators $\hat{p}_0,\hat{p}_1,\hat{p}_2,\hat{p}_3$ with pseudo-fermion operator $c_{i\sigma}$. The first type composite $S=1/2$ states are
\begin{equation}
\begin{aligned}
|\frac{1}{2},\sigma\rangle=\hat{p}_0^{\dagger}c_{\sigma}^{\dagger}|\text{Vac}\rangle.
\end{aligned}
\end{equation}
 Alternatively, we can define $S=1$ eigenstates of vector $\hat{p}$ bosons:
\begin{equation}
\begin{aligned}
 \hat{p}_{1,1}^{\dagger}&=\frac{1}{\sqrt{2}}(\hat{p}_1^{\dagger}-i\hat{p}_2^{\dagger}),\\
 \hat{p}_{1,-1}^{\dagger}&=-\frac{1}{\sqrt{2}}(\hat{p}_1^{\dagger}+i\hat{p}_2^{\dagger}),\\
 \hat{p}_{1,0}^{\dagger}&=-\hat{p}_3^{\dagger},
\end{aligned}
\end{equation}
 and with the Clebsch-Gorden coefficients, we obtain the second type composite $S=1/2$ states
 \begin{equation}
\begin{aligned}
|\frac{1}{2},\frac{1}{2}\rangle&=-\frac{1}{\sqrt{3}} \hat{p}_{1,0}^{\dagger}c_{\uparrow}^{\dagger}|Vac\rangle+\frac{\sqrt{2}}{\sqrt{3}} \hat{p}_{1,1}^{\dagger}c_{\downarrow}^{\dagger}|Vac\rangle,\\
|\frac{1}{2},-\frac{1}{2}\rangle&=\frac{1}{\sqrt{3}} \hat{p}_{1,0}^{\dagger}c_{\downarrow}^{\dagger}|Vac\rangle-\frac{\sqrt{2}}{\sqrt{3}} \hat{p}_{1,-1}^{\dagger}c_{\uparrow}^{\dagger}|Vac\rangle.
\end{aligned}
\end{equation}
Moreover, the combination of the above two types also results in a spin one half particle via defining
  \begin{equation}
\begin{aligned}
|\sigma\rangle =\sum\limits_{\sigma'} \hat{p}_{\sigma \sigma '}^{\dagger} c_{ \sigma '}^{\dagger}|Vac\rangle,
\end{aligned}
\end{equation}
 where $\hat{p}_{\sigma \sigma '}^{\dagger}$ is the matrix elements of
\begin{equation*}
\underline{\hat{p}}^{\dagger} =
  \left[
  \begin{matrix}
   a\hat{p}_0^{\dagger}+b\hat{p}_3^{\dagger}&b(\hat{p}_1^{\dagger}-i\hat{p}_2^{\dagger})  \\
      b(\hat{p}_1^{\dagger}+i\hat{p}_2^{\dagger})&a\hat{p}_0^{\dagger}-b\hat{p}_3^{\dagger}
  \end{matrix}
  \right] .
  \end{equation*}
 From normalization condition, we obtain $a^2+3b^2=1$ and we take $a=b=1/2$, then
  \begin{equation*}
  \begin{aligned}
&\hat{p}_{\sigma \sigma '}^{\dagger}=\frac{1}{2}\sum\limits_{\mu=0}^3 \hat{p}_{\mu}^{\dagger}\tau_{\mu,\sigma\sigma '},  \\
&\hat{p}_{\sigma \sigma '}=\frac{1}{2}\sum\limits_{\mu=0}^3 \hat{p}_{\mu}\tau_{\mu,\sigma\sigma '}.
  \end{aligned}
  \end{equation*}

 To project out the unphysical states in the extended Hilbert space, the following constraints are necessary:
  \begin{equation}
  \begin{aligned}
 & \hat{e}_i^{\dagger}\hat{e}_i+\hat{d}_i^{\dagger}\hat{d}_i+\sum\limits_{\mu}\hat{p}_{i\mu}^{\dagger}\hat{p}_{i\mu}-1=0,\\
 & \sum\limits_{\sigma}c_{i\sigma}^{\dagger}c_{i\sigma}-\sum\limits_{\mu} \hat{p}_{i\mu}^{\dagger} \hat{p}_{i\mu}-2\hat{d}_i^{\dagger }\hat{d}_i=0,\\
&\sum\limits_{\sigma\sigma'}\boldsymbol \tau_{\sigma\sigma'} c_{i\sigma '}^{\dagger}c_{i\sigma}-\hat{p}_{i0} \hat{\boldsymbol p}_i^{\dagger}-\hat{\boldsymbol  p}_i^{\dagger}\hat{p}_{i0}
+i (\hat{\bold p}_i^{\dagger} \times \hat{\bold p}_i)=0.
    \end{aligned}
  \end{equation}
  Alternatively, they can be written as
 \begin{equation}
  \begin{aligned}
&\hat{e}_i^{\dagger}\hat{e}_i+\hat{d}_i^{\dagger}\hat{d}_i+\sum\limits_{\mu}\hat{p}_{i\mu}^{\dagger}\hat{p}_{i\mu}=1,\\
&c_{i\sigma '}^{\dagger}c_{i\sigma}=2\sum\limits_{\sigma_1} \hat{p}_{i\sigma_1 \sigma}^{\dagger} \hat{p}_{i\sigma ' \sigma_1}+\delta_{\sigma \sigma '}\hat{d}_i^{\dagger }\hat{d}_i,\\
    \end{aligned}
    \label{constrain}
  \end{equation}
where the first constraint means that each physical state has one boson, and the second constraint guarantees that correct boson states are attached to the corresponding fermion states.

 One can easily check that with the above constraints and the formulas $[p_{\sigma \sigma '}, p_{\sigma ' \sigma}^{\dagger}]=\frac{1}{2}$ and $p_{\sigma\bar{\sigma}}p_{\sigma\bar{\sigma}}^{\dagger}|\text{Vac}\rangle=0$, only the following four physical states are left:

    \begin{equation*}
  \begin{aligned}
|\uparrow\rangle&= \hat{p}_{\uparrow \uparrow}^{\dagger}c_{\uparrow}^{\dagger}|Vac\rangle+ \hat{p}_{\uparrow \downarrow}^{\dagger}c_{\downarrow}^{\dagger}|Vac\rangle,\\
|\downarrow\rangle&= \hat{p}_{\downarrow \uparrow}^{\dagger}c_{\uparrow}^{\dagger}|Vac\rangle+ \hat{p}_{\downarrow \downarrow}^{\dagger}c_{\downarrow}^{\dagger}|Vac\rangle,\\
|0\rangle&= \hat{e}^{\dagger}|Vac\rangle,\\
|\uparrow\downarrow\rangle&= \hat{d}^{\dagger}c_{\uparrow}^{\dagger}c_{\downarrow}^{\dagger}|Vac\rangle,
    \end{aligned}
  \end{equation*}
where the former two states are singly occupied states, the third is holon and the last is doublon.

  The local $s$ orbital operators $s_{\sigma}$ in our CK model are represented by $s_{\sigma}=\sum_{\sigma '}\hat{z}_{\sigma\sigma '}c_{\sigma '}$, with the matrix $\underline{z}$ defined as:
 \begin{equation}
  \begin{aligned}
\underline{\hat{z}}=\hat{e}^{\dagger}\underline{L} \underline{R}\underline{\hat{p}}+\underline{\hat{\tilde{p}}}^{\dagger}\underline{L}\underline{R}\hat{d},
    \end{aligned}
    \label{definez}
  \end{equation}
  where
\begin{equation}
  \begin{aligned}
&\underline{L}=\left[ (1-\hat{d}^{\dagger}\hat{d})\underline{1}-2\underline{\hat{p}}^{\dagger}\underline{\hat{p}}\right]^{-\frac{1}{2}},\\
&\underline{R}=\left[ (1-\hat{e}^{\dagger}\hat{e})\underline{1}-2\underline{\hat{\tilde {p}}}^{\dagger}\underline{\hat{\tilde{p}}}\right]^{-\frac{1}{2}}.\\
    \end{aligned}
  \end{equation}
Here $\underline{1}$ is a $2\times 2$ identity matrix, and $\underline{\hat{\tilde{p}}}_{\sigma \sigma '}=(\hat{T}\underline{\hat{p}}\hat{T}^{-1})_{\sigma \sigma '}=\sigma \sigma ' \hat{p}_{\bar{\sigma}'\bar{\sigma}}$ is time reversal of $\hat{p}_{\sigma\sigma'}$. The operator $\underline{L}\underline{R}$ in the Eq.~\eqref{definez} equals to identity matrix and acts as a renormalization factor in the mean-field approximation.

 For each $s$ orbital at site $\bold R_i$, a set of above auxiliary operators are induced with index $i$ labeling their sites. In terms of the auxiliary operators and writing the constraints in form of  Lagrange multiplier fields, the CK Hamiltonian takes the form
\begin{equation}
 \begin{aligned}
H=&\sum\limits_{i\sigma} \left[ \sum\limits_{\sigma' \sigma ''}t_{s}^Y \hat{z}_{i\sigma \sigma '}^{\dagger}\hat{z}_{i\pm\hat{Y}\sigma '' \sigma} c_{i\sigma'}^{\dagger} c_{i\pm\hat{Y}\sigma''}-\Delta_s    c_{i\sigma}^{\dagger} c_{i\sigma}+ t_{p}^Xp_{Xi\sigma}^{\dagger} p_{Xi\pm\hat{X}\sigma}\right]+\left[\sum_{\langle ij\rangle \sigma}F(\bold r) \hat{z}_{i\sigma \sigma '}^{\dagger}c_{i\sigma '}^{\dagger}p_{Xj\sigma}\delta_{j,i+\bold r}+\mathrm{H.c.}\right ]\\
&+\sum\limits_{i}    \bigg[ U_s \hat{d}_i^{\dagger}\hat{d}_i +\alpha_{i}(\hat{e}_i^{\dagger}\hat{e}_i+\hat{d}_i^{\dagger}\hat{d}_i+\sum\limits_{\mu}\hat{p}_{i\mu}^{\dagger}\hat{p}_{i\mu}-1)
+{ \beta}_{i0} (\sum\limits_{\sigma}c_{i\sigma}^{\dagger}c_{i\sigma}-\sum\limits_{\mu} \hat{p}_{i\mu}^{\dagger} \hat{p}_{i\mu}-2\hat{d}_i^{\dagger }\hat{d}_i)\\
&+{\boldsymbol \beta}_i\cdot (\sum\limits_{\sigma\sigma'}\boldsymbol \tau_{\sigma\sigma'} c_{i\sigma '}^{\dagger}c_{i\sigma}-\hat{p}_{i0} \hat{\boldsymbol p}_i^{\dagger}-\hat{\boldsymbol  p}_i^{\dagger}\hat{p}_{i0}
+i (\hat{\bold p}_i^{\dagger} \times \hat{\bold p}_i)) \bigg].
  \end{aligned}
  \end{equation}
Here the $U_s \hat{d}_i^{\dagger}\hat{d}_i$ operator equals to the $U_s \hat{n}_{si\uparrow}\hat{n}_{si\downarrow}$ under the constraints Eq.~\eqref{constrain}.

To perform the saddle point approximation, we assume the magnetization is in $X$-$Y$ plane and takes the form of $\bold{M}_i=M\hat{\bold n}_i$, where $\hat{\bold n}_i=(\cos{\phi_i},\sin{\phi_i},0)$, and $\phi_i=\bold Q\cdot \bold R_i$ is site dependent angle. To describe such magnetization in the mean-field theory, we assume that the vector slave-boson order parameter have the same spatial variation as $M_i$, so $\bold {p}_i=p\hat{\bold n}_i$, $\boldsymbol{\beta}_i=\beta \hat{\bold n}_i$. On the other hand, the scalar fields $e_i,p_{0i},\alpha_i, \beta_{0i} $ is assumed to be uniform in real space and $d_i=0$ since $U_s$ is infinitely large. We also assume that all the mean-fields are real numbers.

The matrix $\sqrt{2}\underline{\hat{p}}_i$ which is defined as $\hat{p}_{\sigma \sigma '}=\frac{1}{2}\sum_{\mu=0}^3 \hat{p}_{\mu}\tau_{\mu,\sigma\sigma '}$ has eigenvalues $p_{\nu}=(p_0+\nu p)/\sqrt{2}$ and eigenvectors  $\chi_i^{\nu}=\frac{1}{\sqrt{2}}[\nu e^{-i\phi_i},1]^T$ with $\nu=\pm1$. The matrix $\underline{z}_i$, which is defined as
\begin{equation}
\begin{aligned}
\underline{\hat{z}}_i=\hat{e}_i^{\dagger}\underline{L}_i\underline{R}_i\underline{\hat{p}}_i+\underline{\hat{\tilde{p}}}_i^{\dagger}\underline{L}_i\underline{R}_i\hat{d}_i,
\end{aligned}
\end{equation}
with
\begin{equation}
\begin{aligned}
&\underline{L}_i=\left[ (1-\hat{d}_i^{\dagger}\hat{d}_i)\underline{1}-2\underline{\hat{p}}_i^{\dagger}\underline{\hat{p}}_i\right]^{-\frac{1}{2}},\\
&\underline{R}_i=\left[ (1-\hat{e}_i^{\dagger}\hat{e}_i)\underline{1}-2\underline{\hat{\tilde{p}}}_i^{\dagger}\underline{\hat{\tilde{p}}}_i\right]^{-\frac{1}{2}},\\
\end{aligned}
\end{equation}

can be easily evaluated by writing the $\underline{\hat{p}}_i$ matrix as $\sum_{\nu}p_{\nu}\chi_i^{\nu}\chi_i^{\nu\dagger}/\sqrt{2}$:
\begin{equation*}
\underline{z} _i =
  \left[
  \begin{matrix}
   z_+& z_-e^{-i\phi_i}\\
      z_-e^{i\phi_i}&z_+
  \end{matrix}
  \right],
  \end{equation*}
  where
\begin{equation*}
  \begin{aligned}
&z_{\pm}=ep_+L_+R_-/\sqrt{2}\pm ep_-L_-R_+/\sqrt{2},\\
&L_{\nu}=\left[1-p_{\nu}^2\right]^{-\frac{1}{2}},\\
&R_{\nu}=\left[1-e^2-p_{\nu}^2\right]^{-\frac{1}{2}}.\\
\end{aligned}
\end{equation*}

Now we Fourier transform terms in the mean-field Hamiltonian. Firstly we consider the hopping terms $t_{s,i,j}^Y$ between $s$ orbitals. From the definition $s_{\sigma}=\sum_{\sigma '}\hat{z}_{\sigma\sigma '}c_{\sigma '}$, these terms is represented by the pseudo-fermion operators:
\begin{equation*}
\begin{aligned}
&\sum\limits_{ij\sigma}t_{s,i,j}^Ys_{i\sigma}^{\dagger}s_{j\sigma}\\
=&\sum\limits_{ij\sigma \sigma_1 \sigma_2}t_{s,i,j}^Yz_{i\sigma \sigma_1}^{\dagger}c_{i\sigma_1}^{\dagger}c_{j\sigma_2}z_{j\sigma_2\sigma}\\
=&  \sum\limits_{ij \sigma_1 \sigma_2}\left[\sum\limits_{\sigma} t_{s,i,j}^Yz_{i\sigma \sigma_1}^{\dagger}z_{j\sigma_2\sigma}\right]c_{i\sigma_1}^{\dagger}c_{j\sigma_2}\\
=&\sum\limits_{ij}t_{s,i,j}^Y\left[c_{i\uparrow}^{\dagger}c_{j\uparrow}(z_+^2+z_-^2e^{i(\phi_i-\phi_j)})+c_{i\downarrow}^{\dagger}c_{j\downarrow}(z_+^2+z_-^2e^{i(\phi_j-\phi_i)} ) +c_{i\uparrow}^{\dagger}c_{j\downarrow}(z_+z_-(e^{i\phi_i}+e^{\phi_j})+c_{i\downarrow}^{\dagger}c_{j\uparrow}(z_+z_-(e^{i\phi_i}+e^{\phi_j})            \right].
\end{aligned}
\end{equation*}

The fermion operators in k-space reads:
\begin{equation*}
\begin{aligned}
c_i^{\dagger}=&\sum\limits_{k}e^{ik\cdot R_i}c_k^{\dagger},\\
c_i=&\sum\limits_{k}e^{-ik\cdot R_i}c_k.
\end{aligned}
\end{equation*}
After Fourier transformation, the $t_s^Y$ hopping terms takes the form
\begin{equation*}
\begin{aligned}
  \sum\limits_{ij\sigma}t_{s,i,j}^Ys_{i\sigma}^{\dagger}s_{j\sigma}=&\sum\limits_{\bold k}\biggl[
(z_+^2t_{\bold k}+z_-^2t_{\bold k+\bold Q})c_{\bold k\uparrow}^{\dagger}c_{\bold k\uparrow}+(z_-^2t_{\bold k}+z_+^2t_{\bold k+\bold Q})c_{\bold k+\bold Q\downarrow}^{\dagger}c_{\bold k+\bold Q\downarrow}\\
&+z_+z_-(t_{\bold k+\bold Q}+t_k)c_{\bold k\uparrow}^{\dagger}c_{\bold k+\bold Q\downarrow}+z_+z_-(t_{\bold k+\bold Q}+t_{\bold k})c_{\bold k+\bold Q\downarrow}^{\dagger}c_{\bold k\uparrow}\biggr],
\end{aligned}
\end{equation*}
where the dispersion $t_k$ is defined as $t_{\bold k}=\epsilon_{s\bold k}=2t_s^Y\cos{k_Y}$ in our model.

Secondly we Fourier transform the operators contained in the Lagrangian multiplier into $k$ space:
\begin{equation*}
  \begin{aligned}
\sum\limits_i\boldsymbol \beta_i\cdot \boldsymbol\tau_{\sigma \sigma '}c_{\sigma '}^{\dagger}c_{\sigma}&=\sum\limits_{i}\beta (\cos{\phi_i},\sin{\phi_i},0)\cdot(\tau_x,\tau_y,\tau_z)_{\sigma\sigma'}c_{i\sigma}^{\dagger}c_{i\sigma '}\\
&=\sum\limits_i\beta e^{-i\phi_i}c_{i\downarrow}^{\dagger}c_{i\uparrow}+\beta e^{i\phi_i}c_{i\uparrow}^{\dagger}c_{i\downarrow}\\
&=\sum\limits_{i,\bold k,\bold k'}\left[\beta e^{i(\bold k-\bold Q-\bold k')\cdot \bold R_i}c_{\bold k\downarrow}^{\dagger}c_{\bold k'\uparrow}+\beta e^{i(\bold k+\bold Q-\bold k')\cdot \bold R_i}c_{\bold k\uparrow}^{\dagger}c_{\bold k'\downarrow}\right]\\
&=\sum\limits_{\bold k}\left[\beta c_{\bold k+\bold Q\downarrow}^{\dagger}c_{\bold k\uparrow}+\beta c_{\bold k\uparrow}^{\dagger}c_{\bold k+\bold Q\downarrow}\right].
\end{aligned}
\end{equation*}

Then we Fourier transform the hybridization terms into $k$ space:

\begin{equation*}
  \begin{aligned}
&\sum\limits_{i,\bold r,\sigma}F(\bold r)z_{i\sigma \sigma_1}^{\dagger}s_{i\sigma_1}^{\dagger}p_{X,i+\bold r,\sigma}+\mathrm{H.c.}\\
=&\sum\limits_{i,\bold r}F(\bold r)\left[z_+c_{i\uparrow}^{\dagger}p_{X,i+\bold r,\uparrow}+z_+c_{i\downarrow}^{\dagger}p_{X,i+\bold r,\downarrow}         \right]
+\sum\limits_{i,\bold r}F(\bold r)\left[z_-e^{-i\phi_i}c_{i\downarrow}^{\dagger}p_{X,i+\bold r,\uparrow}+z_-e^{i\phi_i}c_{i\uparrow}^{\dagger}p_{X,i+\bold r\downarrow}         \right]+\mathrm{H.c.}\\
=&\sum\limits_{i,\bold r,\sigma}z_+F(\bold r)c_{i\sigma}^{\dagger}p_{X,i+\bold r,\sigma}\\
&+\sum\limits_{\bold k \bold k'}\sum\limits_{i,\bold r}z_-F(\bold r)e^{i(\bold k-\bold Q)(\bold R_{i}-\bold R_{i+\bold r})}e^{-i(\bold k'-(\bold k-\bold Q))\bold R_{i+\bold r}}c_{\bold k\downarrow}^{\dagger}p_{X\bold k'\uparrow}\\
&+\sum\limits_{\bold k \bold k'}\sum\limits_{i,\bold r}z_-F(\bold r)e^{i(\bold k+\bold Q)(\bold R_i-\bold R_{i+\bold r})}e^{-i(\bold k'-(\bold k+\bold Q))\bold R_{i+\bold r}}c_{\bold k\uparrow}^{\dagger}p_{X\bold k'\downarrow}+\mathrm{H.c.}\\
=&\sum\limits_{\bold k}\biggl[ z_+V_{\bold k}c_{\bold k\uparrow}^{\dagger}p_{X\bold k\uparrow}+z_+V_{\bold k+\bold Q}c_{\bold k+\bold Q\downarrow}^{\dagger}p_{X\bold k+\bold Q\downarrow}+z_-V_{\bold k}c_{\bold k+\bold Q\downarrow}^{\dagger}p_{X\bold k\uparrow}
+z_-V_{\bold k+\bold Q}c_{\bold k\uparrow}^{\dagger}p_{X\bold k+\bold Q\downarrow}\biggr]+\mathrm{H.c.}.
    \end{aligned}
  \end{equation*}

With above results, we have replaced operators $f_{i\sigma}^{\dagger}$ with $c_{i\sigma}^{\dagger}$ and can write the mean-field Hamiltonian in the basis $\bold X_{\bold k}^{\dagger}\equiv (c_{\bold k\uparrow}^{\dagger},c_{\bold k+\bold Q\downarrow}^{\dagger},p_{X\bold k\uparrow}^{\dagger},p_{X\bold k+\bold Q\downarrow}^{\dagger})$ as

\begin{equation}
  \begin{aligned}
  H=&\sum\limits_k \bold X_{\bold k}^{\dagger} \epsilon_k \bold X_{\bold k}+N \Big[-\beta_0(p_0^2+p^2)\\
  &+2\beta p_0p+\alpha(e^2+p^2+p_0^2-1)         \Big ],
\end{aligned}
\end{equation}
   with matrix $\epsilon_k$ defined as:

\begin{equation}
\epsilon_{k}  =
  \left(
  \begin{matrix}
  \epsilon_{s\bold k}^a+\beta_0 &\epsilon_{s\bold k }^c+\beta & z_+V_{\bold k}& z_-V_{\bold k+\bold Q}\\
       \epsilon_{s\bold k }^c+\beta  &\epsilon_{s\bold k}^b+\beta_0& z_-V_{\bold k}&z_+V_{\bold k+\bold Q}\\
        z_+V_{\bold k}^*& z_-V_{\bold k}^*&\epsilon_{p,\bold k}&0\\
      z_-V_{\bold k+\bold Q}^* &z_+V_{\bold k+\bold Q}^*&0&\epsilon_{p,\bold k+\bold Q}
  \end{matrix}
  \right),
  \end{equation}
where $\epsilon_{s\bold k}^a=z_+^2\epsilon_{s\bold k}+z_-^2\epsilon_{s\bold k+\bold Q}-\Delta_s$, $\epsilon_{s\bold k}^b=z_+^2\epsilon_{s\bold k+\bold Q}+z_-^2\epsilon_{s\bold k}-\Delta_s$, $\epsilon_{s\bold k}^c=z_+z_-(\epsilon_{s\bold k+\bold Q}+\epsilon_{s\bold k})$ are $s$ orbital hopping terms.

\noindent

\end{document}